\documentclass[aps,pra,reprint,superscriptaddress,nofootinbib]{revtex4-1}
\usepackage{amsmath}
\usepackage{amssymb}
\usepackage{amsfonts}
\usepackage{mathrsfs}
\usepackage{graphicx}
\usepackage[caption=false]{subfig}
\usepackage{enumitem}
\usepackage[normalem]{ulem}
\usepackage[percent]{overpic}
\usepackage{bm}
\usepackage{rotating}
\usepackage[usenames, dvipsnames]{color}
\usepackage{hyperref}
\usepackage{cancel}
\usepackage{verbatim}
\usepackage{mathtools}
\usepackage{graphicx}
\usepackage{tensor}
\usepackage{verbatim}
\usepackage{wrapfig}
\usepackage{booktabs}

\graphicspath{ {images/} }
\usepackage[caption=false]{subfig}
\usepackage[export]{adjustbox}

\usepackage{braket}
\usepackage[vskip=1em,font=itshape,leftmargin=2em,rightmargin=2em]{quoting}
\newcommand{\ii}{\mathrm{i}}

\renewcommand{\d}{\mathrm{d}}

\newcommand{\be}{\begin{equation}}
\newcommand{\bel}[1]{\begin{equation}\label{#1}}
\newcommand{\ee}{\end{equation}}

\begin{document}
\title{Dimensional reduction of cavities with axial symmetry:\\
A complete analysis of when an optical fiber is approximately one-dimensional}
\author{Daniel Grimmer}
\email{daniel.grimmer@philosophy.ox.ac.uk}
\affiliation{Pembroke College, University of Oxford, Oxford, OX1 1DW, UK}

	\author{Richard Lopp}
	\affiliation{Institute for Quantum Computing, University of Waterloo, Waterloo, ON, N2L 3G1, Canada}
	\affiliation{Department of Applied Mathematics, University of Waterloo, Waterloo, ON, N2L 3G1, Canada}
	\email{rplopp@uwaterloo.ca}

	\author{Eduardo Mart\'{i}n-Mart\'{i}nez}
	\affiliation{Institute for Quantum Computing, University of Waterloo, Waterloo, ON, N2L 3G1, Canada}
	\affiliation{Department of Applied Mathematics, University of Waterloo, Waterloo, ON, N2L 3G1, Canada}
	\affiliation{Perimeter Institute for Theoretical Physics, 31 Caroline St N, Waterloo, ON, N2L 2Y5, Canada}
	\email{emartinmartinez@uwaterloo.ca}

\begin{abstract}
Intuition dictates that a very long, very thin cavity (e.g., a fiber optic cable) could perhaps be modeled as an approximately one dimensional system. In this paper we rigorously explore the validity of such intuition from the perspective of a localized probe coupling to a quantum field inside a cavity (e.g., an atom or an Unruh-DeWitt particle detector in a fiber optic cable). To do so, we introduce the notion of \textit{subfield decomposition} in which a $D+1$ dimensional quantum field in an axially-symmetric cavity can be reduced to an infinite collection of uncoupled, massive $1+1$ dimensional fields. We show that the ability to approximate a higher-dimensional scenario by a $1+1$ dimensional model is equivalent to making a certain change of the probe's shape in the higher-dimensional space. The approximation is justified whenever this change of shape is ``small enough''. In this light, we identify the dynamically relevant norm by which the magnitude of these changes in probe shape ought to be judged. Finally, we explore this approximation in particular setups corresponding to quantum optics and superconducting circuits.
\end{abstract}

\maketitle

\section{Introduction}
There is an intuitive sense in which a long, thin cavity (e.g., a fiber optic cable) can be modeled as an approximately one dimensional system. Approximations along these lines are ubiquitous in the quantum optics literature, see e.g.~\cite{fuentes2011,aida2014,wang2014,kimble2012,liberato, scully2003, scully2019, 2004Natur431162W, PhysRevA.81.062131}. But how exactly does this dimensional reduction work technically? And under exactly what conditions is such an approximation valid? In this paper we answer these questions from the perspective of a probe system which couples locally to a quantum field inside of a cavity. We will model the probe with the Unruh-DeWitt (UDW) model~\cite{PhysRevD.14.870, deWitt}, which has been proven to faithfully capture the main features of the light-matter interaction when the exchange of angular momentum between probe and field does not play a relevant role, and the quantum nature of the center of mass of the probe can be neglected~\cite{Martin-MartinezMOnteroDelRey,Pozas2016,LoppM2021}. The UDW model is more general than those typically used in quantum optics. Indeed, under further approximations, it recovers the typical models used in quantum optics such as spin-boson or Jaynes-Cummings models~\cite{Emma,Pablo,LoppM2021}.

In this paper we will show how a $D+1$ dimensional quantum field inside of a cavity can be mapped (without any approximation) to an infinite collection of massive $1+1$ dimensional quantum fields, which we call subfields. We will discuss this subfield decomposition in sufficient generality to apply it to a wide variety of cavity geometries and boundary conditions. After this, we identify the \textit{dimensional reduction approximation} as the approximation made by ignoring all but one of these subfields. 

It is important to note that since the subfields are generically massive (even if the $D+1$ field is massless), one cannot in general approximate a 3+1 dimensional field in a very long, very thin optical fiber by a~\textit{massless} $1+1$ dimensional field, as is often done. When done properly, and in the regimes where this is possible, the dimensional reduction approximates a long, thin, 3+1 dimensional cavity by a $1+1$ dimensional theory with a mass given by the transverse modes scales~\cite{LoppMPage}.

The goal of this paper is not to explore when the $1+1$ dimensional field should be thought of as massive (as was done in~\cite{LoppMPage}) rather our goal is to explore when the reduction to one (or perhaps a few) of the subfields itself is justified. Given that we are attempting to justify this approximation from the perspective of a localized probe system, the question becomes: under what conditions is only one of the subfields relevant for the probe's evolution? 

As we will discuss, the subfield decomposition tells us the strength with which the probe couples to each of the subfields. These coupling strengths are fixed entirely by the size and shape of the probe in the $D+1$ dimensional description. Thus, the dimensional reduction approximation---which, recall, takes the probe to couple to only one subfield---is equivalent to an approximation on the probe's shape. Thus the question of which subfields are relevant to the probe's dynamics reduces to the question of which changes in the probe's shape will only minimally affect its evolution.

As we will see, the question of whether two shapes are similar enough from the perspective of a probe is nontrivial; it does not generally coincide with what we, as physicists or mathematicians (or even just as people with human eyes) may, \textit{a priori}, consider similar enough. Concretely, we will discuss how the $L^1$ and $L^2$ distances between probe shapes are very ill-suited for this task. Furthermore, we will identify the dynamically relevant norm for the shape comparison by considering how changes in the probe's shape affect its evolution. This new norm captures the way that the probe ``sees'' the space around it. 
 
 

The paper is organized as follows. In Sec.~\ref{LightMatter} we review the Unruh-DeWitt model as a scalar analogue of the electromagnetic dipole coupling, which shall serve as our model for the interaction between light and  a probe within a cavity. In Sec.~\ref{Reduction}, for the case of  Dirichlet boundary conditions, we show how the interaction of a probe with a $D+1$ dimensional theory can be written in terms an infinite number of (massive) subfields, and how the probe couples to each individually. In Sec.~\ref{OtherCases}, we generalize the results for a wide range of boundary conditions and cavity geometries. In Sec.~\ref{Justified} we discuss the relationship between performing a dimensional reduction approximation and changing the probe's shape and size. Finally, in Sec.~\ref{Transitions}, we show the example of a cylindrical cavity when the dimensional reduction can be justified in quantum optics and superconducting circuits regimes. We present our conclusions in Sec.~\ref{Conclusions}.

\section{Probing quantum fields: the Unruh-DeWitt model and the light-matter interaction}\label{LightMatter}
To properly model light-matter interactions one needs to consider the electromagnetic field. This, however, bears difficulties such as the vector nature and gauge aspects of the theory. Here, in turn, we will consider a common simplification of the coupling of the electromagnetic field to matter: the so-called Unruh-DeWitt model (UDW) \cite{PhysRevD.14.870, deWitt}. The vector field will be replaced by a massless scalar field, $\hat\phi$, and matter couples to the field through a monopole moment, $\hat\mu$, which represents the internal degrees of freedom of an Unruh-DeWitt detector. We usually call the detector `the probe' since it allows us to gather properties of the field with local measurements without a need for projective measurements applied directly on the field, which is known to be incompatible with relativity~\cite{sorkin1993impossible,Pipolast}.

It has been shown that this simplified scalar model  captures the relevant features of the light-matter interaction if one neglects the exchange of angular momentum between field and detector \cite{Martin-MartinezMOnteroDelRey, Pozas2016}, and the quantum nature of the center of mass \cite{LoppM2021}. The typical models used in quantum optics for the light-matter interaction (such as Jayne-Cummings or spin-boson) are recovered under non-relativistic approximations of the UDW model~\cite{NichoFaster,Pablo,Emma,LoppM2021}.

Concretely, consider a $D+1$ dimensional real (potentially massive\footnote{Note that we allow this $D+1$ dimensional field to be massive only to increase the generality of our consideration. None of the conclusions or techniques discussed in this paper depend critically on having $M\neq0$.}) non-interacting scalar field of mass $M$, $\hat{\phi}(t,\bm x)$, with \mbox{$D\geq1$}. The free Hamiltonian for such a field is
\begin{align}\label{HamPhi}
\hat{H}_\textsc{f}
&=\frac{1}{2}\int \d^{D}\bm{x} \  c^2\hat{\pi}(t,\bm{x})^2
+\vert\nabla\hat{\phi}(t,\bm{x})\vert^2
+\frac{M^2c^2}{\hbar^2}\hat{\phi}(t,\bm{x})^2,
\end{align}
where $\hat{\pi}(t,\bm{x})$ is the canonically conjugate momentum to $\hat{\phi}(t,\bm{x})$. We consider a probe system which is coupled locally to the field in the interaction picture via the interaction Hamiltonian 
\begin{align}\label{HamInt}
\hat{H}_\textsc{i}(t)
=  g\, \chi(t) \int\d^{D}\bm{x} \  F(\bm{x}) \, \hat{\mu}(t)\otimes\hat{\phi}(t,\bm{x}),
\end{align}
 where $g$ is the coupling strength. $\chi(t)$ and $F(\bm{x})$ are the switching and smearing functions, respectively, controlling when and where the probe couples to the field.

\section{Dimensional reduction of Cavities with axial symmetry}\label{Reduction}
\begin{figure}[t]
	\includegraphics[scale=0.6]{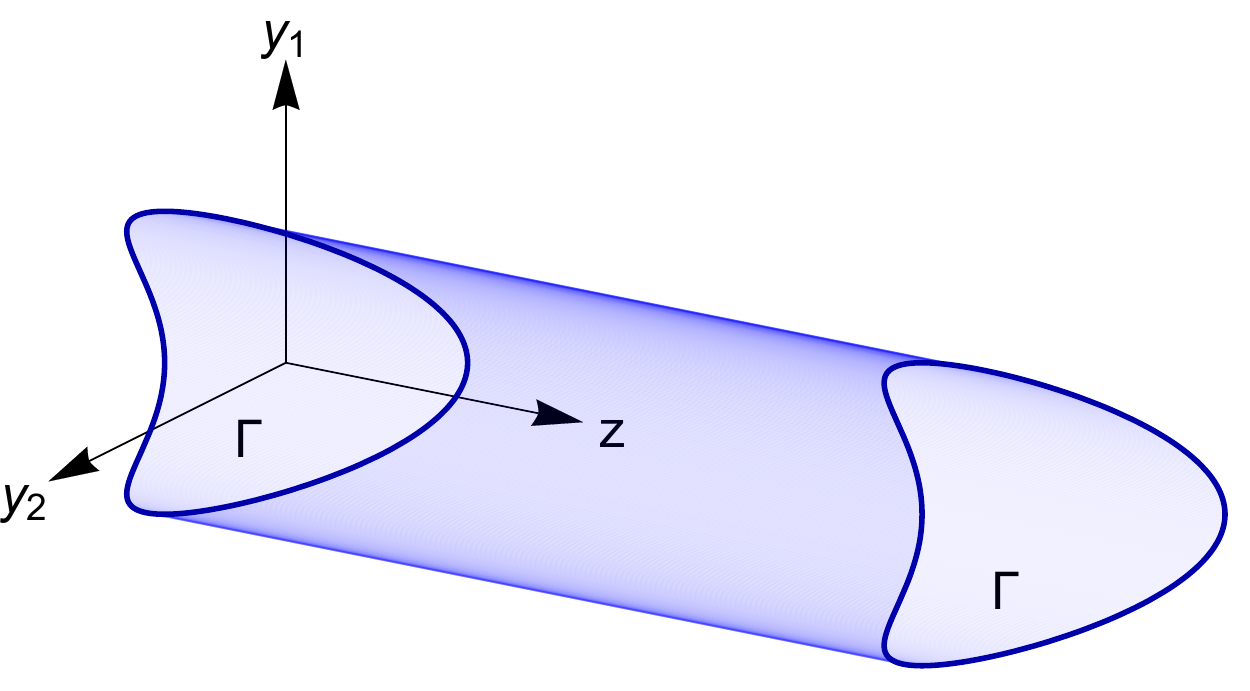}
	\caption{An example of a cavity with axial symmetry.}
	\label{AxialCavity}
\end{figure}

Consider a $D+1$ dimensional real massive free scalar field, $\hat{\phi}(t,x_1,\dots,x_{D-1},x_D)$, with \mbox{$D\geq2$} as discussed in Sec.~\ref{LightMatter}. We are particularly interested in such a field living in a cavity with an axial symmetry with arbitrary cross-section $\Gamma$ (see Fig.~\ref{AxialCavity}). To this end, we partition the spatial dimensions as \mbox{$\bm{x}=(\bm{y},z)$} with  $\bm{y}=(x_1,\dots,x_{D-1})$ and $z=x_D$. We take the cavity to be extended along its axial coordinate, $z$, from $z=0$ to $z=L$. In the transverse coordinates, $\bm{y}$, we take the cavity to have an arbitrary shape defined by $\bm{y}\in\Gamma$ where $\Gamma\subset\mathbb{R}^{D-1}$ is a bounded domain. For instance, if $\Gamma$ defines a triangle then the cavity is a triangular prism. If $\Gamma$ defines a disk, then the cavity is cylindrical.  

We will leave the boundary conditions in the $z$ direction (at $z=0$ and $z=L$) unspecified: e.g. Dirichlet, Neumann, periodic, etc. For sake of introduction, we will in this section take the field to obey Dirichlet boundary conditions in the $\bm{y}$ directions, i.e. $ \hat{\phi}(t,\bm{y},z)=0$ for all $\bm{y}\in\partial\Gamma$. In Sec.~\ref{OtherCases} we will discuss the generalization to other transverse boundary conditions.

The field's free Hamiltonian~\eqref{HamPhi} written in these $(\bm{y},z)$ coordinates is
\begin{align}\label{HamPhiAxial}
\hat{H}_\textsc{f}
=\frac{1}{2}\int_\Gamma \d\bm{y}
\int_0^L\d z \, \Big( 
&c^2\,\hat{\pi}(t,\bm{y},z)^2
+\vert\nabla\hat{\phi}(t,\bm{y},z)\vert^2\\
\nonumber
&+\frac{M^2c^2}{\hbar^2}\hat{\phi}(t,\bm{y},z)^2\Big),
\end{align}
where $\hat{\pi}(t,\bm{y},z)$ and  $\hat{\phi}(t,\bm{y},z)$ satisfy the equal-time canonical commutation relations
\begin{align}
[\hat{\phi}(t,\bm{y}_1,z_1),\hat{\pi}(t,\bm{y_2},z_2)]\!=\ii\hbar\,\hat{\openone}\,\delta(\bm{y}_1-\bm{y}_2)\,\delta(z_1-z_2).
\end{align}
The field-probe interaction Hamiltonian~\eqref{HamInt} reads likewise 
\begin{align}\label{HamIntAxial}
\hat{H}_\textsc{i} 
= g \, \chi(t) \int_\Gamma \d\bm{y}\int_0^L\d z \, F(\bm{y},z) \, \hat{\mu}(t)\otimes\hat{\phi}(t,\bm{y},z).
\end{align}

One may expect that if the cavity is a \textit{very thin fiber}, then one should be able to approximate the interaction of the probe and the $D+1$ dimensional field by a simpler interaction of the probe and an effectively $1+1$ dimensional field. Concretely, let $R$ be a characteristic lengthscale of $\Gamma$. (For instance, if $\Gamma$ is a disk, $R$ could be its radius. If $\Gamma$ defines a polygon, $R$ could be its inradius.) One might expect the dimensional reduction criteria to be $L\gg R$. Certainly, in this regime the cavity \textit{looks} one-dimensional from the outside, but is this what the probe sees from inside?

Our approach to answering this question is as follows: First we will show how the $D+1$ dimensional field in a cavity can be recast (with no approximations) as an infinite collection of uncoupled $1+1$ dimensional fields (what we will call \textit{subfields}). We will then identify how the probe interacts with each of these $1+1$ dimensional subfields. Viewed in this light, the desired dimensional reduction can be identified as making an approximation in which the probe couples to only one (or maybe a few) of these subfields. Perhaps surprisingly, we will see that the dimensional reduction condition is not as simple as $L\gg R$, it also involves the size/shape of the probe as well as the duration of its interaction with the field and the initial field state.

\subsection{Subfield Decomposition}
To map the $D+1$ dimensional quantum field into a set of simpler $1+1$ dimensional fields, we first split the derivative, $\nabla$, in~\eqref{HamPhiAxial} into its axial and transversal components ($\nabla=\partial_z+\nabla_\Gamma$), yielding
\begin{align}
\hat{H}_\textsc{f}
=\frac{1}{2}\int_\Gamma \d\bm{y}\int_0^L& \d z  \, \Big(c^2\hat{\pi}(t,\bm{y},z)^2
+(\partial_z\hat{\phi}(t,\bm{y},z))^2\\
\nonumber
&+\vert\nabla_\Gamma\,\hat{\phi}(t,\bm{y},z)\vert^2
+\frac{M^2c^2}{\hbar^2}\hat{\phi}(t,\bm{y},z)^2\Big).
\end{align}
Integration by parts in the transverse directions, $\bm{y}$, gives
\begin{align}
\hat{H}_\textsc{f}
=\frac{1}{2}\int_\Gamma \d\bm{y}\int_0^L&\d z \, \Big(c^2\hat{\pi}(t,\bm{y},z)^2
+(\partial_z\hat{\phi}(t,\bm{y},z))^2\\
\nonumber
&+\hat{\phi}(t,\bm{y},z)\left(\Delta_\Gamma
+\frac{M^2c^2}{\hbar^2}\right)\hat{\phi}(t,\bm{y},z)\Big),
\end{align}
where $\Delta_\Gamma$ is the Dirichlet Laplacian over $\Gamma$, that is the Laplacian restricted to operating on functions which vanish on $\partial\Gamma$. Note that we have used the boundary condition over $\Gamma$ to remove the boundary term.

We next find the eigenfunctions $\psi_j(\bm{y})$ and eigenvalues  $\lambda_j$ of this transversal Laplacian. These obey 
\begin{align}\label{Eigenstuff}
\Delta_\Gamma\,\psi_j(\bm{y})=-\lambda_j \psi_j(\bm{y}),
\quad
\psi_j(\bm{y})=0\quad\text{for }y\in\partial\Gamma
\end{align}
for integer $j\geq1$. Since $\Gamma$ is bounded we know the Dirichlet Laplacian has an unbounded discrete positive spectrum, \mbox{$0<\lambda_1\leq\lambda_2\leq\lambda_3\leq\dots\to\infty$}, and that its eigenfunctions form an orthonormal basis with respect to the $L^2$ inner product \cite{review-laplacian}.

Several example geometries will be considered in detail in Sec.~\ref{OtherCases}. For now, and to start building intuition, let us focus on the simple case of a rectangular cavity.  Let $\Gamma=[0,L_1]\times[0,L_2]\times\dots\times[0,L_d]$ for a \mbox{$d$-dimensional} rectangular cavity. The eigenfunctions and eigenvalues of the Dirichlet Laplacian are then
\begin{align}
\psi_{n_1,\dots,n_d}(\bm{y})
&=\prod_{k=1}^d \sqrt{\frac{2}{L_k}} \sin\left(\frac{n_k\,\pi\,y_k}{L_k}\right),\\
\lambda_{n_1,\dots,n_d}
&=
\sum_{k=1}^d \left(\frac{n_k\,\pi}{L_k}\right)^2,
\end{align}
for integers $n_k\geq1$. Note that these eigenvalues have multiple indices, $n_1,\dots,n_d$. In order to be converted into the single index form \eqref{Eigenstuff} these eigenvalues would need to be listed and sorted.

We can expand the field operators at each $t$ and $z$ in terms of these eigenfunctions as
\begin{align}
\hat{\phi}(t,\bm{y},z)
&=\sum_{j=1}^\infty 
\hat{\phi}_{j}(t,z)  \,
\psi_j(\bm{y}),\\
\hat{\pi}(t,\bm{y},z)
&=\sum_{j=1}^\infty
\hat{\pi}_{j}(t,z)  \,
\psi_j(\bm{y}),
\end{align}
where
\begin{align}
\hat{\phi}_j(t,z)
&\coloneqq
\int_\Gamma\d\bm{y} \  
\hat{\phi}(t,\bm{y},z) \,
\psi_j(\bm{y}),\\
\hat{\pi}_j(t,z)
&\coloneqq
\int_\Gamma\d\bm{y} \  
\hat{\pi}(t,\bm{y},z) \,
\psi_j(\bm{y}),
\end{align}
are $1+1$ dimensional fields which we will dub ``subfields''. Indeed these subfields obey the equal-time canonical commutation relations
\begin{align}
[\hat{\phi}_i(t,z_1),\hat{\pi}_j(t,z_2)]=\ii\,\hbar\,\hat{\openone}\,\delta(z_1-z_2)\delta_{ij}.
\end{align} 
The field's free  Hamiltonian~\eqref{HamPhiAxial} can be written in terms of the subfields as
\begin{align}\label{HamPhiSubfields}
\hat{H}_\textsc{f}
=\sum_{j=1}^\infty 
\frac{1}{2}\int \d z \ 
& c^2\,\hat{\pi}_j(t,z)^2
+(\partial_z\hat{\phi}_j(t,z))^2\nonumber\\
&+\left(\frac{M^2  c^2}{\hbar^2}+\lambda_j\right)\hat{\phi}_j(t,z)^2.
\end{align}
Note that the subfields are all uncoupled from each other and each have an effective mass $M_j$ given by 
\begin{align}
M_j
=\frac{\hbar}{c}\sqrt{\frac{M^2  c^2}{\hbar^2}+\lambda_j}.
\end{align}
Or equivalently,
\begin{align}
(Mc^2)^2
=(M_j c^2)^2-(\hbar\,c\, k_j)^2,
\quad\text{where }\lambda_j\eqqcolon k_j^2.
\end{align}
Note that even if the original field were massless, $M=0$, the subfields are still effectively massive since \mbox{$\lambda_j>0$} (due to the transverse Dirichlet boundary conditions).  Neglecting this effective mass has been shown to yield incorrect predictions for probes coupling to quantum fields in cavities, especially in relativistic regimes~\cite{LoppMPage}. 

The masses of these subfields can be thought of as being due to a  confinement effect. We note that light inside of an idealized box (small and massless with perfectly reflecting walls) behaves inertially~\cite{Robles_2012}. That is, the light-box behaves in many ways exactly like a massive particle would. In actuality, our field in a cavity is confined (at least in the transverse directions). This confinement does not go away as the cavity becomes ``more one-dimensional'', indeed the field is \textit{more} confined in this limit/regime. As we will discuss more in-depth soon, as the transverse scale of the cavity, $R$, becomes increasingly small, the eigenvalues $\lambda_j$ (and therefore the subfield masses $M_j$) become increasingly large. Said differently, the cavities transverse geometry does not simply ``go away'' in the thin cavity limit, it remains present and is encoded in the subfield masses.


\subsection{The probe's interaction with the subfields}
Now that we have decomposed the $D+1$ dimensional field into these uncoupled massive $1+1$ dimensional subfields, we can investigate how the probe couples to each subfield. Recall that our goal is to identify under what conditions we might be able to approximate the probe's interaction with the full field with an interaction with only  one (or maybe a few) of these $1+1$ dimensional subfields.

The field-probe interaction Hamiltonian~\eqref{HamIntAxial} can be straightforwardly written in terms of the subfields as
\begin{align}\label{HamIntSubfields}
\hat{H}_\textsc{i}
&=\sum_{j=1}^\infty 
g \, \chi(t) \int_0^L \d z \, F_j(z) \, \hat{\mu}(t)\otimes\hat{\phi}_j(t,z),
\end{align}
where
\begin{align}\label{SubfieldsSmearing}
F_j(z) &\coloneqq 
\int_\Gamma\d\bm{y} \  
F(\bm{y},z) \,
\psi_j(\bm{y})
\end{align}
is the probe's effective smearing function for the $j^\text{th}$ subfield. Note that for a generic smearing function $F(\bm{y},z)$ the probe will couple to every subfield to varying degrees. Given this, we may ask: under what conditions is a single (or few) subfield approximation justified? We will consider this question in detail in Sec.~\ref{Justified} but we can already see two hints as to why we might be able to ignore the subfields with very high index, $j$. 

Firstly, if the probe's smearing function is relatively smooth, then we can expect the probe to couple very weakly to subfields with high index $j$; the corresponding eigenfunction $\psi_j(\bm{y})$ will be highly oscillatory and so have little overlap with the relatively smooth $F(\bm{y},z)$. (A slight wrinkle here however is that atomic probes are often modeled as being point-like, i.e., not smooth. More on this in Sec.~\ref{Justified}.)

Secondly, these high-j subfields will have  high eigenvalues $\lambda_j$ and therefore  high effective masses $M_j$. If the subfield's effective mass is large compared to the energy scales associated with the probe-field coupling we expect that the coupling will not provide ``enough energy'' to excite (or absorb energy from) highly massive subfields. This would effectively decouple subfields with large $j$ from the probe. (The intricacy here is that in the $R\to0$ limit where we intuitively expect to get dimensional reduction, \textit{all} of the subfield masses will diverge. In this limit the probe's interaction with every subfield will be ``frozen out''. More on this in Sec.~\ref{Transitions}.)

Before exploring in detail exactly when a single subfield approximation is justified, let us first discuss how this ``subfield decomposition'' can be achieved for more general cavity geometries and boundary conditions.

\section{General Boundary Conditions and Example Geometries}\label{OtherCases}
As we showed in the previous section, one can reduce a $D+1$ dimensional field in an axially symmetric cavity to an infinite collection of uncoupled effectively massive $1+1$ dimensional fields. We demonstrated this for a cavity with rectangular cross-section $\Gamma=[0,L_1]\times[0,L_2]\times\dots\times[0,L_d]$ and Dirichlet boundary conditions on $\partial\Gamma$. However, as we will now discuss this reduction can be done much more generally. 

Indeed, the only properties of the Dirichlet Laplacian that we used are that it has a discrete spectrum and that its eigenfunctions form a complete orthonormal basis. If $\Gamma$ is open, bounded, connected and has a piecewise smooth  boundary (i.e., Lipschitz), then the Neumann Laplacian and Robin Laplacian also have these properties\footnote{Recall that the Dirichlet (Neumann, Robin) Laplacian, is the Laplacian restricted to operating on functions which obey Dirichlet (Neumann Robin) boundary conditions, i.e. acting only on functions which obey these boundary conditions. Functions which obey Dirichlet boundary conditions vanish on the boundary: $f(x)=0$ on $\partial\Gamma$. Neumann boundary conditions are satisfied when the function has a derivative of zero at the boundary in the direction normal to the boundary. That is, Neumann boundary conditions are satisfied when $\partial_n f(x)=0$ on $\partial\Gamma$, where $\partial_n$ is the directional derivative in a direction normal to the boundary. Robin boundary conditions are satisfied when a certain weighted sum of the function value and the normal derivative has a certain value along the boundary. That is, $a\,f(x)+b\,\partial_n f(x)=c$ on $\partial\Gamma$ for some fixed $a$, $b$, and $c$.}~\cite{courant-hilbert,courant-hilbert2}. We could even consider periodic boundary conditions for $\Gamma$. More generally we could take $\Gamma$ to be any compact manifold. This possibility will be discussed further on in this section. This widens the scope of the above subfield decomposition to include a huge number of different transverse geometries and boundary conditions. All that changes case-to-case are the eigenvalues $\lambda_j$ (and so the effective masses $M_j$) and the eigenfunctions $\psi_j(\bm{y})$ (and so the effective smearing functions $F_j(z)$).

One key difference between different boundary conditions is their allowance or disallowance of a  ``constant'' eigenfunction with eigenvalue $\lambda=0$, i.e. a zero mode. For instance, this will always happen in the Neumann case and never in the Dirichlet case. When there is a subfield with $\lambda=0$ and when the $D+1$ dimensional field is massless, $M=0$, then there can be a single massless subfield.

The relationship between the eigenvalues of the Laplacians and the geometry of $\Gamma$ deserves some further comment.
In particular, any eigenvalue of the Robin Laplacian is lower and upper bounded by the corresponding eigenvalues (labeled by $j$) of the Neumann and Dirchlet Laplacian, respectively. This is a consequence of the Courant minmax principle~\cite{courant-hilbert}. It further implies domain monotonicity for Dirichlet boundary conditions (however not for Robin and Neumann)  so that $\lambda_j(\Gamma_1)\geq\lambda_j(\Gamma_2)$ if $\Gamma_1 \subset \Gamma_2$~\cite{review-laplacian}.
Moreover, while the distribution of the eigenvalues $\lambda_j$ is fixed by the shape of $\Gamma$ and the boundary conditions, the reverse is not true. These eigenvalues do not carry complete information about the shape of $\Gamma$ \cite{kac-drum} (one cannot always hear the shape of a drum). However, the works by Kempf et al. prove that the spectrum of the Laplacian does carry a great deal of information about this shape~\cite{kempf2013, kempf2017}. A standard example of the relationship between the spectrum of the Laplacian and the shape of its domain is Weyl's law \cite{Weyl1911,Weyl1912} which tells us that for large $j$ the eigenvalue $\lambda_j$ scales as 
\begin{align}\label{WeylsLaw}
\lambda_j\sim 4\pi^2\left(\frac{j}{V_d \ \vert\Gamma\vert}\right)^{2/d},
\end{align}
where $d=D-1$ is the dimension of the domain $\Gamma$, $V_d$ is the volume of the unit-ball in $\mathbb{R}^d$ and $\vert\Gamma\vert$ is the \mbox{$d$-dimensional} volume of $\Gamma$. 

Note that since \mbox{$\vert\Gamma\vert\sim R^d$} (where we recall $R$ is the characteristic length scale of $\Gamma$) we have that $\lambda_j\sim R^{-2}$ for small $R$. Thus for all $j$ we have that $M_j\to\infty$ as $R\to0$ (unless of course if $\lambda_j=0$ and $M=0$ such that $M_j=0$). Then in the thin cavity limit, every subfield has an infinite effective mass (except for potentially one subfield with $M=0$). If one is to take this limit, it must be handled carefully. In particular, approximating the probe's interaction as being with a single $1+1$ dimensional \textit{massless} field becomes increasingly inappropriate as $R\to0$. 

This close connection between the effective mass of the subfields and the cavity's geometry leads to the exciting possibility of extracting detailed geometric information from measurements of our probe. Given sufficient measurement data from a probe with a known shape, $F(\bm{y},z)$, but in an unknown geometry it is not unreasonable to imagine that we could extract from this data some approximate values for $M_j$ and $F_j(z)$. From these we can approximately determine $\lambda_j$ and $\psi_j(\bm{y})$ which together tell us approximately about the geometry of $\Gamma$ and its boundary conditions.

As an extreme (but hopefully delighting) example consider a cosmology with $d=D-1$ compactified spatial dimensions and one\footnote{Note that although we focus here on only one extended longitudinal dimension, there is essentially no barrier to extending the results of our paper to cases with multiple extended spatial directions.} extended spatial direction. The inhabitants of such a cosmos may be completely unaware of these compactified dimensions. Such inhabitants would likely interpret their particle physics experiments as indicating the existence of a finite collection of $1+1$ dimensional fields with some distribution of masses. Once these inhabitants learn of the compactified dimensions (maybe taking inspiration from string theory) they can reinterpret these $1+1$ dimensional fields as the subfields of a single $D+1$ dimensional field. By noting the scaling of $M_j$ for high $j$ they could (by Weyl's law) determine the number of compactified dimensions and the $d$-dimensional volume of $\Gamma$. By guessing the geometry of $\Gamma$ they would be able to predict the masses of new yet-to-be-discovered subfields (that they may perhaps call strings?).

Connecting back to quantum optics, this tells us that it may be possible to determine similar geometric information about a  fiber optic cable from the response of a probe embedded in it. For instance, one could estimate the cross sectional area of the cable or detect slight imperfections in the cable's shape. 


\subsection{Example Geometries}\label{sec geo}
For reference, we will now list the eigenfunction and eigenvalues of the Laplacian for several simple geometries and boundary conditions.   

As we mentioned in Sec.~\ref{Reduction}, for a $d=D-1$ dimensional rectangle, $\Gamma=[0,L_1]\times[0,L_2]\times\dots\times[0,L_d]$, the eigenfunctions and eigenvalues of the Dirichlet Laplacian are
\begin{align}\label{RectangleDirichlet}
\psi_{n_1,\dots,n_d}(\bm{y})
&=\prod_{k=1}^d \sqrt{\frac{2}{L_k}} \sin\left(\frac{n_k\,\pi\,y_k}{L_k}\right),\\
\lambda_{n_1,\dots,n_d}
&=\sum_{k=1}^{d} \left(\frac{n_k\,\pi}{L_k}\right)^2,
\end{align}
for integers $n_k\geq1$. The eigenfunctions and eigenvalues of the Neumann Laplacian in this geometry are
\begin{align}\label{RectangleNeumann}
\psi_{n_1,\dots,n_d}(\bm{y})
&=\prod_{k=1}^d A(k,n_k) \cos\left(\frac{n_k\,\pi\,y_k}{L_k}\right),\\
\lambda_{n_1,\dots,n_d}
&=\sum_{k=1}^{d} \left(\frac{n_k\,\pi}{L_k}\right)^2,
\end{align}
for integers $n_k\geq0$ and where $A(k,n)$ is $\sqrt{2/L_k}$ for $n\neq0$ and $\sqrt{1/L_k}$ for $n=0$.

For a disk with radius $R$ the eigenfunctions and eigenvalues of the Dirichlet Laplacian are
\begin{align}\label{Cylinder}
\psi_{m\ell}(r,\varphi)
&=\frac{1}{\sqrt{\pi} R J_{m+1}(x_{m\ell})}\exp(\ii m\varphi) \,  J_m(x_{m\ell}\,r/R),
\end{align}
with eigenvalues $\lambda_{m\ell}=(x_{m\ell}/R)^2$ where $J_m$ is the $m$-th Bessel function and $x_{m\ell}$ is the $\ell$-th zero of the $m$-th Bessel function. Note that each of the above expressions (except for $m=0$) gives us two eigenfuctions, namely its real and imaginary parts. The eigenfunctions and eigenvalues of the Neumann Laplacian for a disk can be found in \cite{review-laplacian}.

The eigenvalues of the Dirichlet, Neumann and Robin Laplacians for an equilateral triangle and many other related geometries are known in closed form \cite{EquilateralDirichlet,EquilateralNeumann,EquilateralRobin}. 

Moreover, much is known about how the spectrum of the Laplacian changes under perturbations to the domain.
  In the case of Dirichlet boundary conditions, the perturbed spectrum converges to the original spectrum for a wide variety of deformations~\cite{courant-hilbert, hale, review-laplacian}. The Neumann Laplacian, on the other hand, can be very sensitive to general perturbations~\cite{davies2002, hale} which makes numerical stability difficult. Indeed, on certain bounded domains the Neumann spectrum may not even be discrete~\cite{hempel}.


\section{Subfield Truncation as a Modification of Probe Shape}\label{Justified}

The previous sections have discussed how we can (with no approximation) rewrite the interaction of a probe with a single $D+1$ dimensional field into  the interaction of a probe with an infinite collection of $1+1$ dimensional fields (which we called subfields). Under what conditions can we approximate this situation as a probe interacting with only one (or perhaps a few) subfields?

This approximation can be straightforwardly achieved by truncating the sum over $j$ in the interaction Hamiltonian Eq.~\eqref{HamIntSubfields} to  include only $j\in J$ for some  finite set of positive integers $J$. Crucially, this truncation is equivalent to modifying the probe's smearing function as
\begin{align}\label{FTR}
F(\bm{y},z)=\!\!\sum_{j=1}^\infty F_j(z)\psi_j(\bm{y})
\to
F^{(\textsc{tr})}(\bm{y},z)\coloneqq\!\!\sum_{j\in J} \! F_j(z)\psi_j(\bm{y}).
\end{align}
A probe with such a truncated smearing function does not couple to subfields with $j\notin J$. This also means that we can truncate the field's free Hamiltonian \eqref{HamPhiSubfields} without further approximation. In other words, from the perspective of the detector, the subfields with $j\notin J$ are completely invisible. Therefore if we exclude them from the description of the field,  any prediction on the detector will not be altered.  In summary: the effects of subfield truncation can be understood entirely in terms of modifying the ``shape'' of the probe.

But how well can the truncation of the subfield sum describe the actual physics of a probe in a thin cavity? One may intuit that this truncation will be justified if the corresponding change of the probe's shape, i.e., 
\begin{align}
\Delta F^{(\textsc{tr})}(\bm{y},z)\coloneqq F(\bm{y},z)- F^{(\textsc{tr})}(\bm{y},z),
\end{align}
is ``small''. That is, if $F(\bm{y},z)$ and $F^{(\textsc{tr})}(\bm{y},z)$ are roughly the same. Let us proceed uncritically (without delving into what ``small'' means) temporarily and see what we find.

For a single subfield approximation to work we would need that \mbox{$F(\bm{y},z)\approx F^{(\textsc{tr})}(\bm{y},z)= f(z)\psi_j(\bm{y})$} for some $f(z)$ and some $j$. That is, such an approximation would be justified only if the probe's smearing function $F(\bm{y},z)$ is sufficiently ``near'' to a harmonic mode of the shape $\Gamma$ and ``far'' from all the other harmonic modes. Such a probe would need to have non-negligible spatial support over the whole cavity (since every harmonic mode of $\Gamma$ is supported over the whole cavity). However, smearing functions for realistic probes (e.g., atoms with shapes roughly given by atomic orbitals) look nothing like these harmonic modes, they are far too localized. By this argument it may seem that a single subfield approximation is never justified for realistic atomic probes.

Indeed, in quantum optics, atoms are taken to be extremely localized, often being modeled as point-like with \mbox{$F(\bm{x})\approx\delta(\bm{x}-\bm{x}_0)$} for some $\bm{x}_0$. In this case, the severity of above described issues are drastically increased. No probe can simultaneously be highly localized (be delta-like) and couple to exactly one subfield (be $\psi_j$-like). In order for a point-like approximation and a single-subfield approximation to hold simultaneously the probe's smearing function would have to be simultaneously ``close'' to two very different spatial distributions:  $f(z)\psi_j(\bm{y})$ and $\delta(\bm{x}-\bm{x}_0)$. If we compare these distributions with the $L^2$ norm\footnote{Recall that the $L^2$ norm of a function is the integral of the modulus square of the function. By contrast the $L^1$ norm is the integral of the modulus of the function.}, these functions are infinitely far apart, \mbox{$\vert\vert f(z)\psi_j(\bm{y})-\delta(\bm{x}-\bm{x}_0)\vert\vert_2=\infty$}. Moreover, with respect to the $L^1$ norm we have
\begin{align}
\vert\vert f(z)\psi_j(\bm{y})-\delta(\bm{x}-\bm{x}_0)\vert\vert_1
&=\vert\vert f(z)\psi_j(\bm{y})\vert\vert_1\\
\nonumber
&\quad+\vert\vert\delta(\bm{x}-\bm{x}_0)\vert\vert_1
\end{align}
since $f(z)\psi_j(\bm{y})$ has no ``volume'' over $\bm{x}_0$. That is, these two distributions saturate the triangle inequality; they are as far apart as possible given their finite $L^1$ norms. Surely, the smearing function $F(\bm{x})$ cannot be simultaneously close to two such distant distributions. By this argument, it appears that the point-like approximation and the single subfield approximation are incompatible.

But surely this conclusion is suspect. In quantum optics it is common to take both the point-like approximation and the $1+1$ dimensional approximation and nothing seems to go horribly wrong there. Indeed, as we will now discuss there is something subtly wrong with the above argument; in infinite dimensional vector spaces, the notions of ``near'' and ``far'' require careful qualification. Different norms do give radically different notions of what it means to be ``far''. Above we saw that $f(z)\psi_j(\bm{y})$ and $\delta(\bm{x}-\bm{x}_0)$ are far apart with respects to both the $L^2$ and $L^1$. However, before we started discussing distances in terms of the $L^2$ and $L^1$ norms, we should have asked, ``why are these the relevant norms for the comparison in this particular physical scenario?'' Indeed, what is a very different shape for our eyes (or for our mathematical intuition based on the $L^1$ and $L^2$ norms) may not be that different from the perspective of a detector coupling to the field. 

More concretely, let us provide some quantitative analysis on how well the $L^2$ distance between the probe's full and truncated functions captures the actual change in the probe's dynamics. As a benchmark for the approximation, rather than relying on our own mathematical intuition to guess the appropriate metric, maybe it is more reasonable to simply consult the probe itself how bad the approximation is. To do this, we can compute the probe's excitation and de-excitation probabilities due to its interaction with the field using either $F(\bm{y},z)$  or $F^{(\textsc{tr})}(\bm{y},z)$. If this truncation does not effect these transition probabilities much, then we can say the approximation is good. 

As an example, we take the probe to be a two-level system with free Hamiltonian \mbox{$\hat{H}_\textsc{P}=\hbar\Omega\,\hat{\sigma}_z/2$} coupled to the field via
\begin{align}
\hat{\mu}(t)=e^{\ii\Omega t}\ket{e}\!\bra{g}+e^{-\ii\Omega t}\ket{g}\!\bra{e}. 
\end{align}
That is, $\hat{\mu}(t)$ mediates transitions between the probe's ground state $\ket{g}$ and excited state $\ket{e}$ which have an energy difference $\hbar\Omega$. This is the most common implementation of the Unruh-DeWitt particle detector model \cite{deWitt}. 

The transition probabilities for such a probe are well known (see, e.g., \cite{Louko_2006}). To leading order in interaction strength $g$
\begin{align}\label{Ppm}
P_{\pm}
=g^2\int\!\!\d t\!\!\int\!\!\d\bm{x}\!\!\int\!\!\d t'\!\!\int\!\!\d\bm{x}'
&\Big(\chi(t)\,F(\bm{x})\,\chi(t')\,F(\bm{x'})\\
\nonumber
&\times e^{\mp\ii \Omega (t-t')} \, W(t,\bm{x},t',\bm{x}')\Big),
\end{align}
where $P_+$ is the probability for the probe to transition from the ground to excited state, $P_-$ is the probability for the probe to transition from the excited to the ground state, and where
\begin{align}\label{Wightman}
W(t,\bm{x},t',\bm{x}')    
\coloneqq\text{Tr}(\hat\rho_\textsc{f}\,\hat{\phi}(t,\bm{x})\,\hat{\phi}(t',\bm{x}'))
\end{align}
is the field's Wightman (two-point) function with $\hat{\rho}_\textsc{f}$ being the initial field state. 

To analyze the difference in detector response for the exact smearing and the truncated one, we evaluate the relative difference in the transition probabilities:
\begin{align}\label{FoMProb}
\delta_{P_{\pm}}\coloneqq\frac{\vert P_{\pm}-P_{\pm}^{(\textsc{tr})}\vert}{P_{\pm}}.
\end{align}
This figure of merit is more appropriate to analyze how ``blind'' the detector is to the truncation of the smearing function. Indeed, it is built directly from the difference the truncation causes in the probe's response. It also naturally takes into account the ``contextual details'' of the detector's interaction (that is, the state of the field, and the coupling strength's dependence on time).

\begin{figure}[t]
	\includegraphics[scale=0.9]{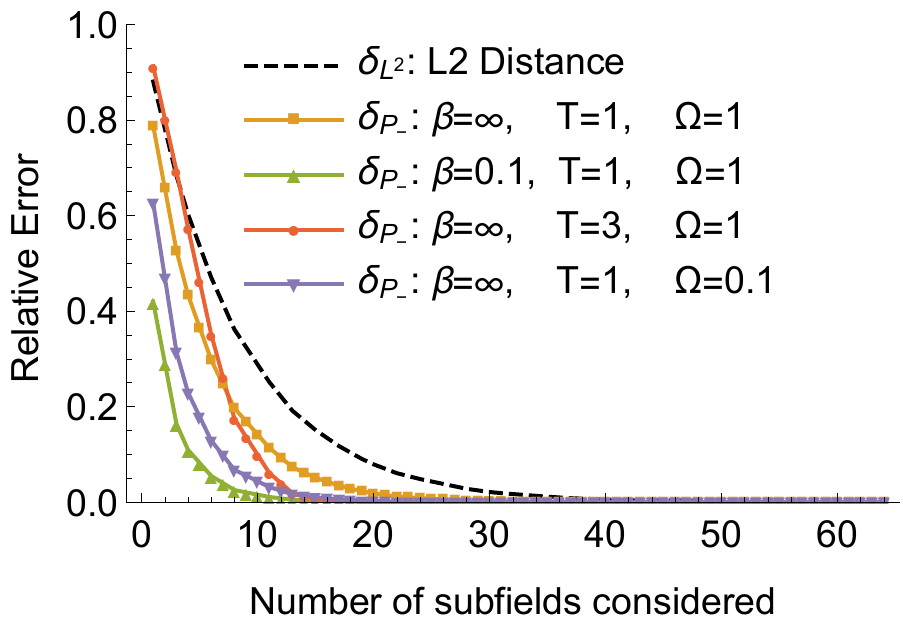}
	\caption{The relative error in the probe's smearing function measured via the $L^2$ distance ($\delta_{L^2}$) is plotted together with the relative error in the  probability of spontaneous emission ($\delta_{P_{-}}$)  as we increase the number of subfields considered. In the $x$-axis the subfields are ordered from lowest to highest effective mass (skipping those which the probe does not couple to out of symmetry, $F_j(z)=0$). Here the probe's smearing function, $F(\bm{y},z)$ is a Gaussian with standard deviation $\sigma=1$ ($\hbar=c=1$) located in the center of the cavity. The probe's switching function, $\chi(t)$, is Gaussian with standard deviation $T$. The cavity has a square cross-section \mbox{$\Gamma=[0,20\sigma]\times[0,20\sigma]$} and length $L=1000\sigma$ with Dirichlet boundary conditions. The field is massless and in a thermal state with inverse temperature $\beta$. Note that we only vary the  switching function, the probe's energy gap, and the field's temperature. We do not change the smearing function or the cavity geometry such that, for all curves in the plot, $F(\bm{y},z)$ and $F^{(\textsc{tr})}(\bm{y},z)$ are identical.}
	\label{RelErrFig}
\end{figure}

In Fig. \ref{RelErrFig} we compare $\delta_{P_{-}}$ with the relative $L^2$ error caused by the truncation
\begin{align}\label{FoML2}
\delta_{L^2}=\frac{\vert\vert F(\bm{y},z)-F^{(\textsc{tr})}(\bm{y},z)\vert\vert_2^2}{\vert\vert F(\bm{y},z)\vert\vert_2^2}
\end{align}
for a range of different field temperatures, coupling times and probe gaps. The relative error in $L^2$ distance (dashed) is insensitive to all of these parameters. However, as Fig.~\ref{RelErrFig} shows, the relative error in $P_-$ is highly sensitive to the parameters. The number of subfields needed to accurately capture probe's spontaneous emission is highly context dependent.

For instance, consider the case (yellow in Fig.~\ref{RelErrFig}) of $\hbar c\beta/\sigma=\infty$, $ c\,T/\sigma=1$, and $\sigma\Omega/c=1$. In this case,  approximately 30 subfields are needed for convergence of the spontaneous emission probability. Moreover, an error of $\approx90\%$ is made by the single subfield approximation. If we increase the field temperature to $\hbar c\beta/\sigma=0.1$ (green) then only 10 subfields are needed for convergence of $P_-$ and the single subfield approximation gives only a $40\%$ error. In all cases, the $L^2$ overestimates the number of field modes needed for convergence and overestimates the errors of a few-subfield approximation. This reveals that while the $L^2$ norm of $\Delta F^{(\textsc{tr})}(\bm{y},z)$ being small may be sufficient to justify truncating the number of subfields, it is apparently not necessary. 

Ultimately, the $L^2$ distance fails because it does not account for the context dependence of the validity of the few subfield approximation. Despite the failure of the $L^1$ and $L^2$ norms to predict when a single (or few) subfield approximation may be valid, we maintain our earlier claim that this approximation can be understood entirely in terms of modifying the shape of the probe. There must be some better context-sensitive norm with which we can appropriately judge the smallness of $\Delta F^{(\textsc{tr})}(\bm{y},z)$. Indeed, from Eq.~\eqref{Ppm} we can identify two such norms:  
\begin{align}\label{NormK}
\frac{P_\pm}{g^2}=\vert\vert F\vert\vert_\pm^2
\coloneqq\int\!\!\d\bm{x}\int\!\!\d\bm{x}'\, K_\pm(\bm{x},\bm{x}') \, F(\bm{x}) \,F(\bm{x}'),
\end{align}
with  kernels $K_\pm(\bm{x},\bm{x}')$ given by
\begin{align}\label{KernelK}
K_\pm(\bm{x},\bm{x}')
\coloneqq\!\!\int\!\!\d t\!\! \int\!\!\d t' \,
\chi(t)\,\chi(t')\,
e^{\mp \ii \Omega (t-t')}
W(t,\bm{x},t',\bm{x}').
\end{align}
These norms naturally take into account the probe's switching function $\chi(t)$ and the state of the field through the field's Wightman function, $W(t,\bm{x},t',\bm{x}')$. Moreover, one can check that we have
\begin{align}\label{RelDiff}
\frac{\vert\vert F(\bm{y},z)-F^{(\textsc{tr})}(\bm{y},z)\vert\vert_\pm^2}{\vert\vert F(\bm{y},z)\vert\vert_\pm^2}
=\frac{\vert P_{\pm}-P_{\pm}^{(\textsc{tr})}\vert}{P_{\pm}}.
\end{align}
That is, the relative error in these new norms is exactly the relative error in the transition probabilities. This confirms that, ultimately, the way that the probe ``sees'' space (i.e., changes in shape) is through the $\vert\vert F\vert\vert_\pm$ norms rather than trough the $L^1$ or $L^2$ norms. This framing resolves the above raised issue: How can $F(x)$ be near to both $f(z)\psi_j(\bm{y})$ and $\delta(\bm{x}-\bm{x}_0)$ if they are so far apart?  The answer is that $f(z)\psi_j(\bm{y})$ and $\delta(\bm{x}-\bm{x}_0)$ are not so far apart with respect to the norms which are actually relevant for the dynamics.

We can conclude that we have to consider the change of predictions of physical observables, as compared to norms that are insensitive to the physical process, if we want to evaluate the physical impact of truncating the number of subfields.
In the next section, for a concrete example of a cavity geometry, we will see in more detail under what conditions the relative difference~\eqref{RelDiff} of transition probabilities is small.

\section{Dimensional Reduction for a Cylindrical Cavity}\label{Transitions}
Let us further understand under what conditions the $1+1$ dimensional reduction is an adequate approximation through an example.  Consider a cylindrical cavity of radius $R$ and length $L$ with Dirichlet boundary conditions.


 The solution to the $3+1$ dimensional Klein-Gordon equation yields the following mode decomposition of the quantized scalar field,
\begin{align}
	\hat{\phi}(t,r,\varphi,z)= \sum_{\substack{m=0 \\ n,\ell=1}}^{\infty}\left(	u_{m \ell n } \,\hat{a}_{m \ell n } + 	u^*_{m \ell n }\, \hat{a}^{\dagger}_{m \ell n }\right),
	\end{align}
	where the field modes satisfying Dirichlet boundary conditions are
	\begin{align}
	\nonumber 
	u_{m  \ell n}(t,r,\varphi,z)&=A_{m  \ell n}  e^{\ii m \varphi}   e^{ -\ii \omega t } \sin\left(\frac{n \pi}{L} z\right) J_m\left(\frac{x_{m \ell}}{R} r\right), \label{modes} \\
	A_{m \ell n}&=c^{1/2}\left(R \sqrt{L \pi \omega } J_{m+1}(x_{m \ell}) \right)^{-1},\\
	\omega&=c \sqrt{\frac{x^2_{m \ell}}{R^2}+ \frac{n^2 \pi^2}{L^2}},\label{energ}
	\end{align}
	where (see Sec.~\ref{sec geo}) $x_{m l}$ is the $\ell$-th zero of the $m$-th Bessel function of the first kind, $J_m$.
 Note that  the field modes are orthonormalized with respect to the Klein-Gordon inner product. 
 
We assume a Gaussian detector  spatial profile in all directions with central localisation at $z=L/2$, $r=0$ of the form\footnote{Note that assuming the localization of the detector to be much smaller than the cavity dimensions, we can extend the integrals involving the Gaussians to the whole space.}
 \begin{align}
     F(r,\varphi,z)= \frac{1}{\sqrt{2 \pi  \sigma ^2}^3}\exp \left(-\frac{r^2}{2 \sigma ^2}\right) \exp \left(-\frac{\left(z-\frac{L}{2}\right)^2}{2 \sigma ^2}\right)\label{F gauss}
 \end{align}
 which is $L^1$-normalized for a detector strongly localized inside the cavity.
 
Following the procedure laid out in Section~\ref{Reduction}, we can rewrite this scenario as one in which the probe interacts with an infinite collection of $1+1$ dimensional fields, which we call subfields. The transverse profiles of these subfields are given by \eqref{Cylinder}. Note that in this geometry the subfields are labeled by both $m$ and $\ell$. These subfields have effective masses, $M_{m\ell} \coloneqq \frac{\hbar}{c} \sqrt{\lambda_{m\ell}}=\frac{\hbar x_{m\ell}}{c R}$, and mode decompositions,
\begin{align}
	\hat{\phi}_{m\ell}(z,t)= \sum_{ n=1}^{\infty}\left(	\tilde u_{m\ell,n }(t,z) \hat{a}_{n } + 	\tilde u^*_{m\ell,n}(t,z) \hat{a}^{\dagger}_{ n }\right),
	\end{align}
 where 
 \begin{align}
    \tilde u_{m\ell,n}(z,t)= \sqrt{\frac{c}{\tilde \omega L }} e^{-\ii \tilde \omega t} \sin\left(\frac{n \pi}{L} z\right)
\end{align}
and 
 \begin{align}
 \tilde \omega=c\sqrt{\left(\frac{M_{m\ell} c}{\hbar}\right)^2 + \left(\frac{  n \pi}{L}\right)^2} \label{effec energy}.
 \end{align}

Following the procedure around Eq.~\eqref{SubfieldsSmearing}, we can determine how strongly the probe couples to each of these subfields. We find that the probe's smearing function for the $m\ell$ subfield is 
 \begin{align}
     F_{0\,\ell}(z)= \frac{1}{\sqrt{2} \pi  \sigma R J_1(x_{0\ell})} e^{-\frac{\sigma ^2 x_{0\ell}^2}{2 R ^2}}e^{-\frac{\left(z-\frac{L}{2}\right)^2}{2 \sigma ^2}} 
 \end{align}
 for $m=0$ and $F_{m\ell}=0$ for $m\neq0$. Indeed, since our probe is placed on the axis of symmetry of the cavity, the probe does not couple to any of the subfields with $m\neq0$. Later in this section we shall relax the condition that the probe is in the cavity's center.

Consequently, the $3+1$ dimensional theory decomposes into an infinite number of $1+1$ dimensional massive theories as
\begin{align}
\hat{H}_\textsc{i}
&=\sum_{\ell=1}^\infty 
g \, \chi(t) \int_0^L \d z \, F_{0\ell}(z) \, \hat{\mu}(t)\otimes\hat{\phi}_{0\ell}(z,t)
 \label{Fdecomp}
\end{align}

One may expect that truncating the infinite sum could yield a good approximation to the probe dynamics. In particular, the case analyzed in~\cite{LoppMPage} is the special case where the sum is truncated to one subfield, i.e., the interaction is dimensionally reduced to a single $1+1$ dimensional theory. Let us generalize these results by studying the speed at which the transition probability of the probe converges as we consider the effect of more summands in Eq.~\eqref{Fdecomp}.

The number of excitations $N_{\ell n}$ in modes $(\ell,n)$ of the  $3+1$ dimensional model to leading order in perturbation theory is for a Gaussian and a sudden box switching function, respectively, given by (analogous to the one in Appendix B.1 of \cite{LoppMPage})
\begin{align}
	    N_{\ell n}=&\frac{g^2}{\hbar^2}   \left|\int_\mathbb{R}  \d t \chi(t) e^{\pm \ii  \Omega t  } \int_{\mathbb{R}^3} \d^3 x F(r,z)  u_{0\ell, n}^*(r,\varphi,z,t) \right|^2\nonumber\\
	    =&\frac{g^2 c}{\hbar^2} \left|\int_\mathbb{R}  \d t \chi(t) e^{\ii (\pm  \Omega +\tilde\omega)t }  \int_\mathbb{R} \d z  \frac{F_{0,\ell}(z)}{\sqrt{L \omega}}\sin\left(\frac{n \pi}{L} z\right) \right|^2 \nonumber\\
    =&\frac{g^2 c}{\hbar^2 L\omega \pi R^2 J_1(x_{0\ell})^2} e^{-\left(\frac{\pi  n \sigma }{L}\right)^2} e^{-\sigma ^2 M_{0\ell}^2 c^2/\hbar^2} \sin^2\left(\frac{n \pi}{2}\right)  \nonumber\\
    &\times \begin{cases} 
       2\pi T^2 e^{-T^2 (\pm \Omega +\tilde\omega)^2}, & \chi(t)= \exp(-t^2/(2T^2)) \\
     2\frac{ 1-\cos (T (\Omega \pm \tilde\omega))}{(\Omega \pm\omega)^2}, & \chi(t)=\theta(t)-\theta(t-T) 
   \end{cases},    
\label{number}
    	    \end{align}
where  for the $+$ sign the initial state of the detector is the ground state and with the $-$ sign the detector is initially excited.
Then the  vacuum excitation probability $P_+$ and the spontaneous emission probability $P_-$ are obtained via~\cite{LoppMPage}
\begin{align}
   P_{\pm} =\sum_{\ell n=1}^\infty N_{\ell n}=\sum_{\ell=1}^\infty \tilde P_{\pm}^\ell \frac{ e^{-\sigma ^2 M_{0\ell}^2 c^2/\hbar^2}}{\pi R^2 J_1(x_{0\ell})^2},\label{prob}
\end{align}
where the $1+1$ dimensional transition probabilities are defined as 
\begin{align}
    \tilde P_{\pm}^\ell=&\sum_{n=1}^\infty \frac{g^2 c}{\hbar^2 L\tilde{\omega}  } e^{-\left(\frac{\pi  n \sigma }{L}\right)^2}  \sin^2\left(\frac{n \pi}{2}\right)        \nonumber\\
    &\times \begin{cases} 
      2 \pi T^2 e^{-T^2 (\pm \Omega +\tilde{\omega})^2}, & \chi(t)= \exp(-t^2/(2T^2)) \\
     2\frac{ 1-\cos (T (\Omega \pm \tilde{\omega}))}{(\Omega \pm\tilde{\omega})^2}, & \chi(t)=\theta(t)-\theta(t-T) 
   \end{cases}.\label{P cases}
\end{align}
Therefore the $3+1$ transition probability of a detector coupled to a single  massless scalar field is recast as the infinite sum of $1+1$ dimensional subfields with effective mass $M_\ell=\hbar x_{0\ell}/(c R)$ for the corresponding $\ell$.
	\begin{figure*}[p]


			

\centering
\includegraphics[scale=.1]{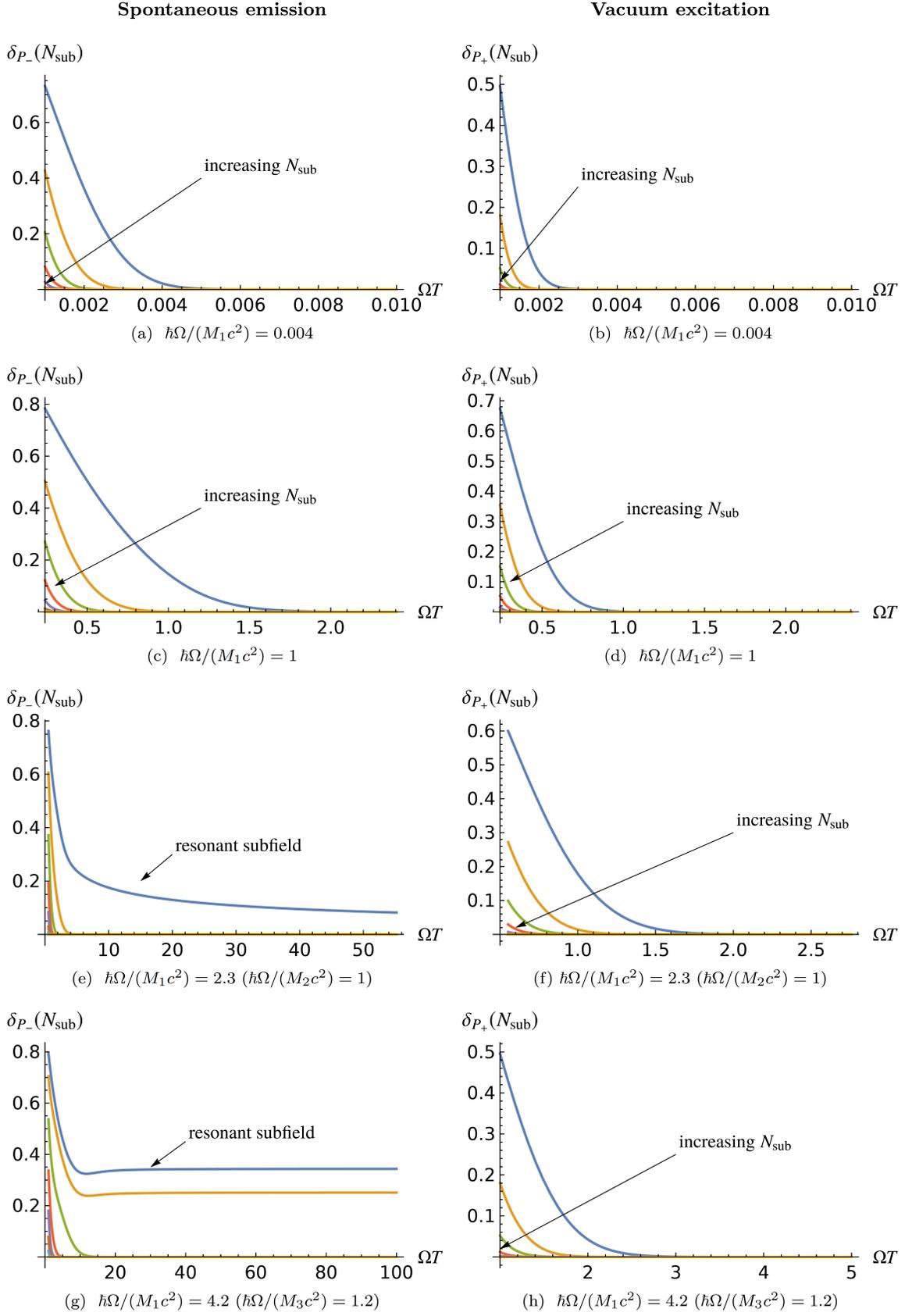}
		\caption{Relative difference $\delta_{P_\pm}(N_\text{sub})$ as a function of interaction time in terms of $\Omega T$ for Gaussian switching. Parameters are $L/R=10^3$ and  (a)--(f) $\sigma/R=10^{-2}$; (g, h) $\sigma/R=10^{-6}$. Most-Resonant subfield in (c,d) is $\ell=1$; in (e,f) $\ell=2$; in (g,h) $\ell=3$; in (a,b) all subfields are off-resonant. Parameters in (a,b) simulate a super-conducting cavity; (g,h) an optical cavity. For vacuum excitation, the single-field approximation will be sufficient for sufficiently long times, depending on the factor $\hbar\Omega/(M_1 c^2)$. In the case of spontaneous emission, the resonant mode will not suffice even in the long-time regime. Here, it is necessary to add the low-mass fields below the mass of the resonance mode. In (e,g) we first added the the resonance modes and subsequently added the fields from $\ell=1$ with increasing $\ell$ to compute $\delta_l$.}
		\label{trans prob gauss}
	\end{figure*}

As the dimensional reduction is usually justified for one spatial dimension being much larger than the remaining dimensions, we will assume $  R\ll L$. Furthermore we assume  $\sigma/R \ll 1$, i.e. the detector is localized far away from the cavity boundaries. 
Two things can be stated already: Firstly, as previously shown in~\cite{LoppMPage}, the factors in Eq.~\eqref{prob} multiplying the $1+1$ dimensional contributions to the probabilities are non-negligible: it would be wrong to just model the long cavity starting from a $1+1$ dimensional theory naively without doing the dimensional reduction. Note also that the dimensions for $g$ are different when starting in one spatial dimension as opposed to the three dimensional case.
Secondly, the frequencies Eq.~\eqref{effec energy} become $\tilde\omega\approx M_{0\ell}c^2/\hbar$ when  $R\ll L$ for small enough $n$. For those large masses it will be very energetically costly to excite the fields and this suggests that the high mass subfields (as compared to the other scales of the problem) will be frozen out and not contribute to the detector dynamics. We will see how this is the case below.

Considering the case of Gaussian switching first, Eq.~\eqref{number} together with the assumption $R\ll L$ imply that the decay of contributions of subfields is governed by 
\begin{equation}
e^{-\frac{\sigma ^2 x_{0\ell}^2}{ R ^2}}e^{-T^2 (\pm \Omega +\tilde\omega)^2}\sim e^{-\sigma ^2 M_{0\ell}^2 c^2/\hbar^2}e^{-T^2 (\pm \Omega + M_{0\ell}c^2/\hbar)^2}\label{gauss_factor}
\end{equation}
for $cT/R\gtrsim 1$, i.e., exponentially suppressed with the effective mass of the subfields\footnote{This strong suppression is a consequence of the smooth Gaussian switching. As we will see, the suppression will not be as strong for any other switching as it goes with the Fourier transform of the switching function.}. This suppression is lower bounded by the spontaneous emission scenario. Thus, we should expect that few subfields will be required for convergence, and that for vacuum excitations convergence sets in more quickly. However, when $cT/R \ll 1$, fast convergence in the number of subfields is only possible if Eq.~\eqref{gauss_factor} is negligible for all but a small set of subfields. For the case of spontanous emission (the minus sign in front of $\Omega$) this happens only if $T \Omega \gg \sigma/R$. In general, we expect thus the convergence in the spontaneous emission case to be slower than the vacuum excitation one.

To quantify the required number $N_\text{sub}$ of subfields, i.e. how many $1+1$ dimensional fields we need to consider for sufficient accuracy in the detector dynamics, we consider the relative difference between the exact transition probability  and a truncated version with only $N_\text{sub}$ summands as per Eq.~\eqref{FoMProb}. In this case, the relative difference can be written as
\begin{align}
    \delta_{P_\pm}(N_\text{sub})=1-\frac{1}{ P_\pm}\sum_{\ell=1}^{N_\text{sub}} \tilde P_{\pm}^{\ell} \frac{ e^{-\sigma ^2 M_{0\ell}^2 c^2/\hbar^2}}{\pi R^2 J_1(x_{0\ell})^2}.
    \label{deltadelta}
\end{align}
In Fig.~\ref{trans prob gauss} we plot the relative difference as a function of $\Omega T$ for different truncations of the sum in Eq.~\eqref{deltadelta} and for different parameter configurations. In particular we consider numerical values exemplary of optical and superconducting cavities. We see that for vacuum excitations, as the evolution time becomes longer, the truncated sum of 1+1D terms approximates the exact calculation better. Indeed, for vacuum excitation processes, even a single-subfield approximation is valid for long times. That can be explained with \eqref{gauss_factor} which suppresses higher order summands strongly with increasing $\ell$. For spontaneous emission, however, the number of subfields is largely governed by being in resonance, i.e. $(\Omega-\tilde\omega)^2$ being close to zero. The determining parameter is $\hbar\Omega/(M_{01} c^2)$, i.e. the ratio of the detector's transition energy to the mass of the least massive subfield.
We find that for increasing values of  $\hbar \Omega/(M_{01} c^2)$ the convergence for a fixed number of subfields becomes slower in time. In fact, even in the long-time regime one or a few subfields will not in general suffice for spontaneous emission probabilities if  $\hbar\Omega$ is large since more subfields will be close to resonance.

Let us consider now the Gaussian switching case with a pointlike probe
\begin{align}
   F^\delta(r,z)= \frac{1}{2\pi r}\delta(r) \delta(z-L/2)
\end{align}
such that $ F^\delta(r,z)$ is $L^1$-normalized. Note that, unlike in~\cite{LoppMPage}---which considered sudden switching--- we keep the Gaussian switching for this pointlike limit.

The number of excitations can be obtained from~\eqref{number}:
\begin{align}
    N_{\ell n}^\delta=\lim_{\sigma \rightarrow 0}  N_{\ell n}.
\end{align}
Therefore, the suppression factor in Eq.~\eqref{gauss_factor} becomes
\begin{equation}
 e^{-T^2 (\pm \Omega + M_{0\ell}c^2/\hbar)^2},\label{dirac factor}
\end{equation}
i.e. there is no longer a suppression factor due to the detector size. Nonetheless, since we chose $\sigma=10^{-2}R$ in Fig.~\ref{trans prob gauss}, the dominant contribution to the decay of \eqref{gauss_factor} came from the remaining exponential in~\eqref{dirac factor}---at least in the long time regime when $\hbar\Omega \geq M_{01} c^2$. Therefore, we do not expect any qualitative differences in the number of required subfields for long times when comparing to the spatially extended case.

\begin{figure}[!ht]
		\centering
\includegraphics[scale=.8,valign=t]{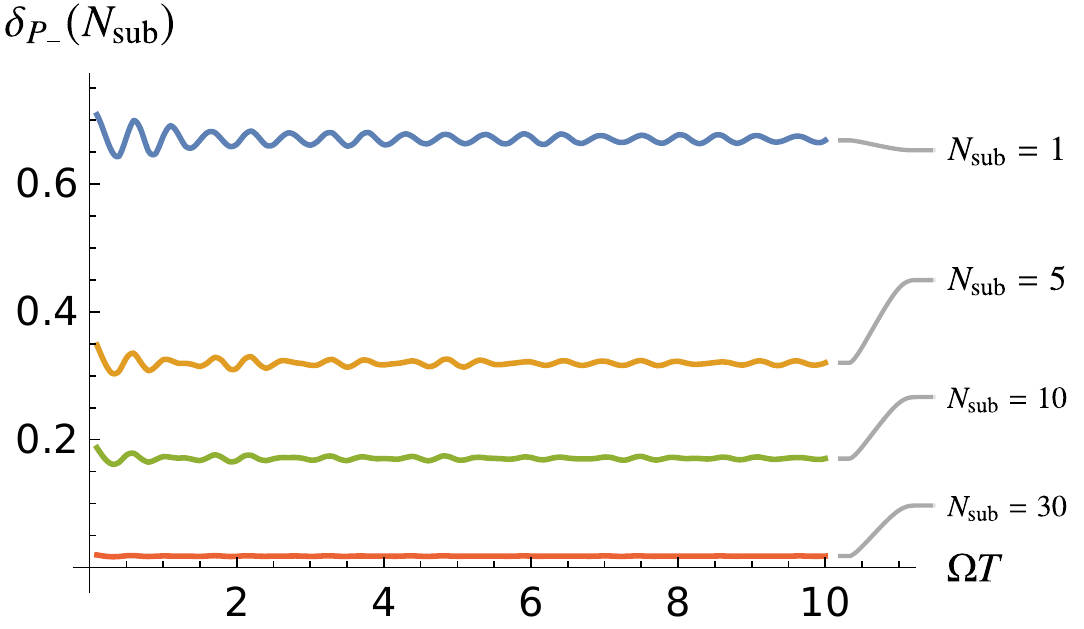}
		\caption{Relative difference $\delta_{P_\pm}(N_\text{sub})$ as a function of interaction time in terms of $\Omega T$ for sudden switching and spontaneous emission. Parameters are $L/R=10^3$, $\sigma/R=10^{-2}$. $\hbar\Omega/(M_{01}c^2)=0.004$. As compared to the Gaussian switching, many more subfields are needed to reduce the relative difference. Note that since we consider $\hbar\Omega\ll M_{01}c^2$, emission and excitation probabilities are near-identical and so we only plotted one transition process. }
		\label{trans prob sudden}
	\end{figure}

Finally, let us now examine the relative difference in the case of a Gaussian smearing but a sudden switching function (as opposed to the adiabatic Gaussian switching above, see Eq.~\eqref{P cases}).
As seen in Fig.~\ref{trans prob sudden}, the transition probability presents oscillatory behavior as a function of  $\Omega T$. Therefore we expect that generally more subfields will be required for a true representation of the transition probability as compared to the Gaussian switching scenario. We find that, already far off-resonance, one needs more subfields as compared to the adiabatic switching case. We also see that the relative difference does not approach zero as a function of $\Omega T$ for a fixed number of subfields, highlighting how the suddenness of the switching prevents a few-subfield approximation even in the long-time regime.

So far we have solely considered detectors that were positioned in the cavity's center,  canceling the contribution from any subfields with $m\neq 0$. We will now study the influence on the convergence of the subfield sum if we relax this assumption.
Before that, let us look at the case where the detector is still centred on the axis of symmetry but not centered around $z=L/2$. 	If the detector is centred  around $z=z_0\in (0,L)$ while still assuming that $z_0 \gg \sigma$ and $L-z_0 \gg \sigma$, then we have to carry out the following replacement in Eq.~\eqref{number}
\begin{align}
   \sin^2\left(\frac{n \pi}{2}\right)   \longrightarrow \sin^2\left(\frac{n z_0 \pi}{L}\right), 
\end{align} i.e. in general the even modes in $n$ will not vanish. Nonetheless, the convergence in the number of subfields is not affected.  
If, however, we position the detector outside of the axis of symmetry of the cavity, for example,  without loss of generality $(r,\varphi)=(r_0,0)$, the spatial smearing \eqref{F gauss} will read 
\begin{align}
     F(r,\varphi,z)= \frac{e^{-\left(z-\frac{L}{2}\right)^2/2 \sigma ^2}}{\sqrt{2 \pi  \sigma ^2}^3}\exp\!\left(-\frac{r^2+r_0^2-2 r_0 r \cos \varphi}{2 \sigma ^2}\right)  .
 \end{align}
 We then have that the transition probability is obtained by Eq.~\eqref{prob} but making the following substitution in the number of excitations in Eq.~\eqref{number}:
 \begin{align}
      e^{-\frac{\sigma ^2 x_{0\ell}^2}{ R ^2}} \to e^{-\frac{r_0^{2}}{ \sigma^{2}}} \left| \int_0^\infty \d r r\frac{e^{-\frac{r^{2}}{2 \sigma^{2}}}}{\sigma^2}   I_m\left( \frac{r_0 r}{\sigma^{2}}\right) J_m\left( \frac{ x_{ml}}{R}r\right)\right|^2,
 \end{align}
 where $I_m$ is the modified Bessel function of the first kind, and the integral is generally non-zero for $m>0$. The convergence in the subfield sum with $m>0$ is subtle and depends on the specific parameters. In general, more than just the leading $m=0$ subfield will be required. Hence it is reasonable to say that the analysis of particle detectors in the absence of axial of symmetry cannot be dimensionally reduced to the coupling with a few 1+1D subfields in general.
 

\subsection*{Particularizing for experimentally meaningful scales}
Through adequate choices of parameters we are able to examine general experimental setups such as superconducting and optical cavities. In particular, if we assume for a typical superconducting cavity a radius of $R=1$ mm and a detector modelling a superconducting qubit of size $\sigma=10\,\mu$m and energy gap $\Omega=10$ GHz (see, e.g., ~\cite{Emma}), the following relations hold:  $\Omega \sigma=10^{-4}$, $\Omega R=10^{-2}$, $\sigma/R=10^{-2}$. Similarly for an optical cavity of radius of $R=0.1$ mm and a detector modelling an atom of size $\sigma=0.1$ nm and energy gap $\Omega=100$ THz, we get $\Omega\sigma=10^{-5}$, $\Omega R=10$, $\sigma/R=10^{-6}$. The scales corresponding to these two experimentally relevant settings are included in the sets of parameters analyzed in Figs.~\ref{trans prob gauss} and~\ref{trans prob sudden} for illustration.

\section{Conclusions and outlook}\label{Conclusions}
In this paper we show how, for the light-matter interaction, one can decompose a $D+1$ dimensional quantum field inside a cavity into an infinite collection of $1+1$ dimensional quantum fields, which we call subfields. We have discussed this subfield decomposition in sufficient generality to apply to a wide variety of cavity geometries and boundary conditions. It is important to note that this subfield decomposition is exact, not an approximation. Using the subfield decomposition that we developed we were able to identify the proper dimensional reduction approximation as the approximation made by ignoring all but one of these subfields. 

One first thing that we clarify is that a naive reduction of a very long, very thin cavity (the so-called the idealized optical fiber limit in previous literature~\cite{LoppMPage}) to a massless 1+1 dimensional field in a cavity is not acceptable in most regimes. This is important because this kind of intuitive (but inaccurate) dimensional reduction has arguably been commonplace in the body of literature on the light-matter interaction. 

One benefit of viewing dimensional reduction in this way is that we can now access a gradation of approximations by considering different numbers of $1+1$ dimensional subfields. This perspective also casts light on which features of the cavity's geometry survive the dimensional reduction and how these features shape the effective $1+1$ dimensional subfields; A triangular cavity and a cylindrical one remain distinguishable in the dimensional reduction limit. In particular, each subfield has an effective mass which encodes information about the cavity's transverse geometry. It can be a grave error to treat these effective $1+1$ dimensional fields as being massless.

Once we made this subfield decomposition, we then investigated exactly how a localized probe system couples to each of these subfields given how it is coupled to the $D+1$ dimensional field. In particular, we have shown that the strength with which the probe couples to each of these subfields is fixed entirely by the size and shape of the probe in the $D+1$ dimensional description. Thus, the dimensional reduction approximation (which takes the probe to couple to only one subfield) can be understood entirely in terms of making an approximation on the probe's shape. Concretely, there are certain special probe shapes for which the probe literally does couple to only one subfield; if we can approximate the probe's actual shape with one of these special shapes, the dimensional reduction will follow exactly. Thus the question becomes: when is this approximation of the probe's shape justified?

As we have seen, this question is actually rather less trivial than it may sound: for the probe to couple to only one subfield, its transverse profile must be (very near to) one of the harmonic modes of the cavity's transverse geometry. These harmonic modes have support over the full width of the cavity. But this is not the shape we typically associate with realistic probes (e.g., atoms). Indeed, realistic probes are typically modeled as being point-like, with a distribution (very near to) a delta-function. These two shapes (harmonic modes and delta-functions) are intuitively very different. However, can the probe's shape be very near to both of them in some meaningful sense?

As we have discussed, a preconception that a delta function is `very different' from a cavity transverse mode stems from the incorrect assumption that the $L^2$ or $L^1$ norms are the relevant measures of the difference of shape. Recall that our goal is to understand when a dimensional reduction approximation is justified from the perspective of the localized probe, not the human perception. That is, we are interested in the conditions under which only one of the subfields are relevant for the probe's evolution.

Roughly, any given subfield might be irrelevant to the probe for one of three reasons: 1) this subfield has a high effective mass (due to the cavity's transverse shape), or 2) it does not couple strongly to the probe (due to the probe's size and shape), or 3) the time profile of the interaction strength suppresses the coupling to some of the subfields. For the dimensional reduction approximation to be justified, these three possibilities must conspire to allow us to eliminate all but one subfield from our consideration. That is, in order for us to justify a dimensional reduction approximation we must think carefully about not only the cavity's shape, but also the shape of the probe within that cavity and its switching function (its ``shape in time''). The relevant norm for judging how far apart a 1+1 dimensional reduction is from the exact model must be sensitive to all of these contextual factors. We have derived such a context-sensitive norm, namely Eq.~\eqref{NormK}, by considering how changes in the probe's shape affect its transition probabilities. This norm captures the way that the probe ``sees'' changes in shape. According to this new norm, a single harmonic transverse mode and the delta function are, in fact, not so far apart in many cases.

Thus to summarize, introducing the notion of a  subfield decomposition we have identified the dimensional reduction approximation as corresponding to a certain change of the probe's shape (i.e., its smearing function). Moreover, we have identified the physically relevant norm by which these changes in probe shape ought to be judged. To explore concretely when this approximation is justified  we have provided a numerical study with typical setups in optical cavities and superconducting circuits in Sec.~\ref{Transitions}.

We believe the tools in this paper can be adopted in quantum optics as a quantifiable standard by which dimensional reduction approximations are justified. Moreover, we believe that the connection between a cavity's transverse geometry and the effective mass of the subfields deserves further study/application. For instance, in detecting the shape (or defects in the shape) of optical cavities: suppose that one has developed a protocol for estimating the masses of $1+1$ dimensional fields from measurements of localized probes coupled to these fields. This exact same protocol applied to localized probes in a $3+1$ dimensional cavity, for instance, would yield detailed information about the cavity's transverse shape. 

\acknowledgments

  The authors would like to thank Achim Kempf for helpful discussions. EMM acknowledges support through the Discovery Grant Program of the Natural Sciences and Engineering Research Council of Canada (NSERC). EMM also acknowledges support of his Ontario Early Researcher award. 



\bibliography{references}

\begin{thebibliography}{36}%
\makeatletter
\providecommand \@ifxundefined [1]{%
 \@ifx{#1\undefined}
}%
\providecommand \@ifnum [1]{%
 \ifnum #1\expandafter \@firstoftwo
 \else \expandafter \@secondoftwo
 \fi
}%
\providecommand \@ifx [1]{%
 \ifx #1\expandafter \@firstoftwo
 \else \expandafter \@secondoftwo
 \fi
}%
\providecommand \natexlab [1]{#1}%
\providecommand \enquote  [1]{``#1''}%
\providecommand \bibnamefont  [1]{#1}%
\providecommand \bibfnamefont [1]{#1}%
\providecommand \citenamefont [1]{#1}%
\providecommand \href@noop [0]{\@secondoftwo}%
\providecommand \href [0]{\begingroup \@sanitize@url \@href}%
\providecommand \@href[1]{\@@startlink{#1}\@@href}%
\providecommand \@@href[1]{\endgroup#1\@@endlink}%
\providecommand \@sanitize@url [0]{\catcode `\\12\catcode `\$12\catcode
  `\&12\catcode `\#12\catcode `\^12\catcode `\_12\catcode `\%12\relax}%
\providecommand \@@startlink[1]{}%
\providecommand \@@endlink[0]{}%
\providecommand \url  [0]{\begingroup\@sanitize@url \@url }%
\providecommand \@url [1]{\endgroup\@href {#1}{\urlprefix }}%
\providecommand \urlprefix  [0]{URL }%
\providecommand \Eprint [0]{\href }%
\providecommand \doibase [0]{http://dx.doi.org/}%
\providecommand \selectlanguage [0]{\@gobble}%
\providecommand \bibinfo  [0]{\@secondoftwo}%
\providecommand \bibfield  [0]{\@secondoftwo}%
\providecommand \translation [1]{[#1]}%
\providecommand \BibitemOpen [0]{}%
\providecommand \bibitemStop [0]{}%
\providecommand \bibitemNoStop [0]{.\EOS\space}%
\providecommand \EOS [0]{\spacefactor3000\relax}%
\providecommand \BibitemShut  [1]{\csname bibitem#1\endcsname}%
\let\auto@bib@innerbib\@empty
\bibitem [{\citenamefont {Dragan}\ \emph {et~al.}(2011)\citenamefont {Dragan},
  \citenamefont {Fuentes},\ and\ \citenamefont {Louko}}]{fuentes2011}%
  \BibitemOpen
  \bibfield  {author} {\bibinfo {author} {\bibfnamefont {A.}~\bibnamefont
  {Dragan}}, \bibinfo {author} {\bibfnamefont {I.}~\bibnamefont {Fuentes}}, \
  and\ \bibinfo {author} {\bibfnamefont {J.}~\bibnamefont {Louko}},\ }\href
  {\doibase 10.1103/PhysRevD.83.085020} {\bibfield  {journal} {\bibinfo
  {journal} {Phys. Rev. D}\ }\textbf {\bibinfo {volume} {83}},\ \bibinfo
  {pages} {085020} (\bibinfo {year} {2011})}\BibitemShut {NoStop}%
\bibitem [{\citenamefont {Ahmadzadegan}\ \emph {et~al.}(2014)\citenamefont
  {Ahmadzadegan}, \citenamefont {Mann},\ and\ \citenamefont
  {Mart\'{\i}n-Mart\'{\i}nez}}]{aida2014}%
  \BibitemOpen
  \bibfield  {author} {\bibinfo {author} {\bibfnamefont {A.}~\bibnamefont
  {Ahmadzadegan}}, \bibinfo {author} {\bibfnamefont {R.~B.}\ \bibnamefont
  {Mann}}, \ and\ \bibinfo {author} {\bibfnamefont {E.}~\bibnamefont
  {Mart\'{\i}n-Mart\'{\i}nez}},\ }\href {\doibase 10.1103/PhysRevA.90.062107}
  {\bibfield  {journal} {\bibinfo  {journal} {Phys. Rev. A}\ }\textbf {\bibinfo
  {volume} {90}},\ \bibinfo {pages} {062107} (\bibinfo {year}
  {2014})}\BibitemShut {NoStop}%
\bibitem [{\citenamefont {Zhu}\ \emph {et~al.}(2014)\citenamefont {Zhu},
  \citenamefont {Wang},\ and\ \citenamefont {Zhou}}]{wang2014}%
  \BibitemOpen
  \bibfield  {author} {\bibinfo {author} {\bibfnamefont {W.}~\bibnamefont
  {Zhu}}, \bibinfo {author} {\bibfnamefont {Z.~H.}\ \bibnamefont {Wang}}, \
  and\ \bibinfo {author} {\bibfnamefont {D.~L.}\ \bibnamefont {Zhou}},\ }\href
  {\doibase 10.1103/PhysRevA.90.043828} {\bibfield  {journal} {\bibinfo
  {journal} {Phys. Rev. A}\ }\textbf {\bibinfo {volume} {90}},\ \bibinfo
  {pages} {043828} (\bibinfo {year} {2014})}\BibitemShut {NoStop}%
\bibitem [{\citenamefont {Chang}\ \emph {et~al.}(2012)\citenamefont {Chang},
  \citenamefont {Jiang}, \citenamefont {Gorshkov},\ and\ \citenamefont
  {Kimble}}]{kimble2012}%
  \BibitemOpen
  \bibfield  {author} {\bibinfo {author} {\bibfnamefont {D.~E.}\ \bibnamefont
  {Chang}}, \bibinfo {author} {\bibfnamefont {L.}~\bibnamefont {Jiang}},
  \bibinfo {author} {\bibfnamefont {A.~V.}\ \bibnamefont {Gorshkov}}, \ and\
  \bibinfo {author} {\bibfnamefont {H.~J.}\ \bibnamefont {Kimble}},\ }\href
  {http://stacks.iop.org/1367-2630/14/i=6/a=063003} {\bibfield  {journal}
  {\bibinfo  {journal} {New J. Phys.}\ }\textbf {\bibinfo {volume} {14}},\
  \bibinfo {pages} {063003} (\bibinfo {year} {2012})}\BibitemShut {NoStop}%
\bibitem [{\citenamefont {S{\'a}nchez~Mu{\~n}oz}\ \emph
  {et~al.}(2018)\citenamefont {S{\'a}nchez~Mu{\~n}oz}, \citenamefont {Nori},\
  and\ \citenamefont {De~Liberato}}]{liberato}%
  \BibitemOpen
  \bibfield  {author} {\bibinfo {author} {\bibfnamefont {C.}~\bibnamefont
  {S{\'a}nchez~Mu{\~n}oz}}, \bibinfo {author} {\bibfnamefont {F.}~\bibnamefont
  {Nori}}, \ and\ \bibinfo {author} {\bibfnamefont {S.}~\bibnamefont
  {De~Liberato}},\ }\href {\doibase 10.1038/s41467-018-04339-w} {\bibfield
  {journal} {\bibinfo  {journal} {Nat. Commun.}\ }\textbf {\bibinfo {volume}
  {9}},\ \bibinfo {pages} {1924} (\bibinfo {year} {2018})}\BibitemShut
  {NoStop}%
\bibitem [{\citenamefont {Scully}\ \emph {et~al.}(2003)\citenamefont {Scully},
  \citenamefont {Kocharovsky}, \citenamefont {Belyanin}, \citenamefont {Fry},\
  and\ \citenamefont {Capasso}}]{scully2003}%
  \BibitemOpen
  \bibfield  {author} {\bibinfo {author} {\bibfnamefont {M.~O.}\ \bibnamefont
  {Scully}}, \bibinfo {author} {\bibfnamefont {V.~V.}\ \bibnamefont
  {Kocharovsky}}, \bibinfo {author} {\bibfnamefont {A.}~\bibnamefont
  {Belyanin}}, \bibinfo {author} {\bibfnamefont {E.}~\bibnamefont {Fry}}, \
  and\ \bibinfo {author} {\bibfnamefont {F.}~\bibnamefont {Capasso}},\ }\href
  {\doibase 10.1103/PhysRevLett.91.243004} {\bibfield  {journal} {\bibinfo
  {journal} {Phys. Rev. Lett.}\ }\textbf {\bibinfo {volume} {91}},\ \bibinfo
  {pages} {243004} (\bibinfo {year} {2003})}\BibitemShut {NoStop}%
\bibitem [{\citenamefont {Ben-Benjamin}\ \emph {et~al.}(2019)\citenamefont
  {Ben-Benjamin}, \citenamefont {Scully}, \citenamefont {Fulling},
  \citenamefont {Lee}, \citenamefont {Page}, \citenamefont {Svidzinsky},
  \citenamefont {Zubairy}, \citenamefont {Duff}, \citenamefont {Glauber},
  \citenamefont {Schleich},\ and\ \citenamefont {Unruh}}]{scully2019}%
  \BibitemOpen
  \bibfield  {author} {\bibinfo {author} {\bibfnamefont {J.~S.}\ \bibnamefont
  {Ben-Benjamin}}, \bibinfo {author} {\bibfnamefont {M.~O.}\ \bibnamefont
  {Scully}}, \bibinfo {author} {\bibfnamefont {S.~A.}\ \bibnamefont {Fulling}},
  \bibinfo {author} {\bibfnamefont {D.~M.}\ \bibnamefont {Lee}}, \bibinfo
  {author} {\bibfnamefont {D.~N.}\ \bibnamefont {Page}}, \bibinfo {author}
  {\bibfnamefont {A.~A.}\ \bibnamefont {Svidzinsky}}, \bibinfo {author}
  {\bibfnamefont {M.~S.}\ \bibnamefont {Zubairy}}, \bibinfo {author}
  {\bibfnamefont {M.~J.}\ \bibnamefont {Duff}}, \bibinfo {author}
  {\bibfnamefont {R.}~\bibnamefont {Glauber}}, \bibinfo {author} {\bibfnamefont
  {W.~P.}\ \bibnamefont {Schleich}}, \ and\ \bibinfo {author} {\bibfnamefont
  {W.~G.}\ \bibnamefont {Unruh}},\ }\href {\doibase 10.1142/S0217751X19410057}
  {\bibfield  {journal} {\bibinfo  {journal} {Int. J. Mod. Phys. A}\ }\textbf
  {\bibinfo {volume} {34}},\ \bibinfo {pages} {1941005} (\bibinfo {year}
  {2019})}\BibitemShut {NoStop}%
\bibitem [{\citenamefont {{Wallraff}}\ \emph {et~al.}(2004)\citenamefont
  {{Wallraff}}, \citenamefont {{Schuster}}, \citenamefont {{Blais}},
  \citenamefont {{Frunzio}}, \citenamefont {{Huang}}, \citenamefont {{Majer}},
  \citenamefont {{Kumar}}, \citenamefont {{Girvin}},\ and\ \citenamefont
  {{Schoelkopf}}}]{2004Natur431162W}%
  \BibitemOpen
  \bibfield  {author} {\bibinfo {author} {\bibfnamefont {A.}~\bibnamefont
  {{Wallraff}}}, \bibinfo {author} {\bibfnamefont {D.~I.}\ \bibnamefont
  {{Schuster}}}, \bibinfo {author} {\bibfnamefont {A.}~\bibnamefont {{Blais}}},
  \bibinfo {author} {\bibfnamefont {L.}~\bibnamefont {{Frunzio}}}, \bibinfo
  {author} {\bibfnamefont {R.-S.}\ \bibnamefont {{Huang}}}, \bibinfo {author}
  {\bibfnamefont {J.}~\bibnamefont {{Majer}}}, \bibinfo {author} {\bibfnamefont
  {S.}~\bibnamefont {{Kumar}}}, \bibinfo {author} {\bibfnamefont {S.~M.}\
  \bibnamefont {{Girvin}}}, \ and\ \bibinfo {author} {\bibfnamefont {R.~J.}\
  \bibnamefont {{Schoelkopf}}},\ }\href {\doibase 10.1038/nature02851}
  {\bibfield  {journal} {\bibinfo  {journal} {Nature}\ }\textbf {\bibinfo
  {volume} {431}},\ \bibinfo {pages} {162} (\bibinfo {year}
  {2004})}\BibitemShut {NoStop}%
\bibitem [{\citenamefont {Lizuain}\ \emph {et~al.}(2010)\citenamefont
  {Lizuain}, \citenamefont {Casanova}, \citenamefont {Garc\'{\i}a-Ripoll},
  \citenamefont {Muga},\ and\ \citenamefont {Solano}}]{PhysRevA.81.062131}%
  \BibitemOpen
  \bibfield  {author} {\bibinfo {author} {\bibfnamefont {I.}~\bibnamefont
  {Lizuain}}, \bibinfo {author} {\bibfnamefont {J.}~\bibnamefont {Casanova}},
  \bibinfo {author} {\bibfnamefont {J.~J.}\ \bibnamefont {Garc\'{\i}a-Ripoll}},
  \bibinfo {author} {\bibfnamefont {J.~G.}\ \bibnamefont {Muga}}, \ and\
  \bibinfo {author} {\bibfnamefont {E.}~\bibnamefont {Solano}},\ }\href
  {\doibase 10.1103/PhysRevA.81.062131} {\bibfield  {journal} {\bibinfo
  {journal} {Phys. Rev. A}\ }\textbf {\bibinfo {volume} {81}},\ \bibinfo
  {pages} {062131} (\bibinfo {year} {2010})}\BibitemShut {NoStop}%
\bibitem [{\citenamefont {Unruh}(1976)}]{PhysRevD.14.870}%
  \BibitemOpen
  \bibfield  {author} {\bibinfo {author} {\bibfnamefont {W.~G.}\ \bibnamefont
  {Unruh}},\ }\href {\doibase 10.1103/PhysRevD.14.870} {\bibfield  {journal}
  {\bibinfo  {journal} {Phys. Rev. D}\ }\textbf {\bibinfo {volume} {14}},\
  \bibinfo {pages} {870} (\bibinfo {year} {1976})}\BibitemShut {NoStop}%
\bibitem [{\citenamefont {DeWitt}(1979)}]{deWitt}%
  \BibitemOpen
  \bibfield  {author} {\bibinfo {author} {\bibfnamefont {B.}~\bibnamefont
  {DeWitt}},\ }\href@noop {} {\emph {\bibinfo {title} {General Relativity: an
  Einstein Centenary Survey}}}\ (\bibinfo  {publisher} {Cambridge University
  Press, edited by S. W. Hawking and W. Israel},\ \bibinfo {year}
  {1979})\BibitemShut {NoStop}%
\bibitem [{\citenamefont {Martín-Martínez}\ \emph {et~al.}(2013)\citenamefont
  {Martín-Martínez}, \citenamefont {Montero},\ and\ \citenamefont {del
  Rey}}]{Martin-MartinezMOnteroDelRey}%
  \BibitemOpen
  \bibfield  {author} {\bibinfo {author} {\bibfnamefont {E.}~\bibnamefont
  {Martín-Martínez}}, \bibinfo {author} {\bibfnamefont {M.}~\bibnamefont
  {Montero}}, \ and\ \bibinfo {author} {\bibfnamefont {M.}~\bibnamefont {del
  Rey}},\ }\href {\doibase 10.1103/PhysRevD.87.064038} {\bibfield  {journal}
  {\bibinfo  {journal} {Phys. Rev. D}\ }\textbf {\bibinfo {volume} {87}},\
  \bibinfo {pages} {064038} (\bibinfo {year} {2013})}\BibitemShut {NoStop}%
\bibitem [{\citenamefont {Pozas-Kerstjens}\ and\ \citenamefont
  {Martín-Martínez}(2016)}]{Pozas2016}%
  \BibitemOpen
  \bibfield  {author} {\bibinfo {author} {\bibfnamefont {A.}~\bibnamefont
  {Pozas-Kerstjens}}\ and\ \bibinfo {author} {\bibfnamefont {E.}~\bibnamefont
  {Martín-Martínez}},\ }\href {\doibase 10.1103/PhysRevD.94.064074}
  {\bibfield  {journal} {\bibinfo  {journal} {Phys. Rev. D}\ }\textbf {\bibinfo
  {volume} {94}},\ \bibinfo {pages} {064074} (\bibinfo {year}
  {2016})}\BibitemShut {NoStop}%
\bibitem [{\citenamefont {Lopp}\ and\ \citenamefont
  {Mart\'{\i}n-Mart\'{\i}nez}(2021)}]{LoppM2021}%
  \BibitemOpen
  \bibfield  {author} {\bibinfo {author} {\bibfnamefont {R.}~\bibnamefont
  {Lopp}}\ and\ \bibinfo {author} {\bibfnamefont {E.}~\bibnamefont
  {Mart\'{\i}n-Mart\'{\i}nez}},\ }\href {\doibase 10.1103/PhysRevA.103.013703}
  {\bibfield  {journal} {\bibinfo  {journal} {Phys. Rev. A}\ }\textbf {\bibinfo
  {volume} {103}},\ \bibinfo {pages} {013703} (\bibinfo {year}
  {2021})}\BibitemShut {NoStop}%
\bibitem [{\citenamefont {McKay}\ \emph {et~al.}(2017)\citenamefont {McKay},
  \citenamefont {Lupascu},\ and\ \citenamefont
  {Mart\'{\i}n-Mart\'{\i}nez}}]{Emma}%
  \BibitemOpen
  \bibfield  {author} {\bibinfo {author} {\bibfnamefont {E.}~\bibnamefont
  {McKay}}, \bibinfo {author} {\bibfnamefont {A.}~\bibnamefont {Lupascu}}, \
  and\ \bibinfo {author} {\bibfnamefont {E.}~\bibnamefont
  {Mart\'{\i}n-Mart\'{\i}nez}},\ }\href {\doibase 10.1103/PhysRevA.96.052325}
  {\bibfield  {journal} {\bibinfo  {journal} {Phys. Rev. A}\ }\textbf {\bibinfo
  {volume} {96}},\ \bibinfo {pages} {052325} (\bibinfo {year}
  {2017})}\BibitemShut {NoStop}%
\bibitem [{\citenamefont {Mart\'{\i}n-Mart\'{\i}nez}\ and\ \citenamefont
  {Rodriguez-Lopez}(2018)}]{Pablo}%
  \BibitemOpen
  \bibfield  {author} {\bibinfo {author} {\bibfnamefont {E.}~\bibnamefont
  {Mart\'{\i}n-Mart\'{\i}nez}}\ and\ \bibinfo {author} {\bibfnamefont
  {P.}~\bibnamefont {Rodriguez-Lopez}},\ }\href {\doibase
  10.1103/PhysRevD.97.105026} {\bibfield  {journal} {\bibinfo  {journal} {Phys.
  Rev. D}\ }\textbf {\bibinfo {volume} {97}},\ \bibinfo {pages} {105026}
  (\bibinfo {year} {2018})}\BibitemShut {NoStop}%
\bibitem [{\citenamefont {Lopp}\ \emph {et~al.}(2018)\citenamefont {Lopp},
  \citenamefont {Mart{\'{\i}}n-Mart{\'{\i}}nez},\ and\ \citenamefont
  {Page}}]{LoppMPage}%
  \BibitemOpen
  \bibfield  {author} {\bibinfo {author} {\bibfnamefont {R.}~\bibnamefont
  {Lopp}}, \bibinfo {author} {\bibfnamefont {E.}~\bibnamefont
  {Mart{\'{\i}}n-Mart{\'{\i}}nez}}, \ and\ \bibinfo {author} {\bibfnamefont
  {D.~N.}\ \bibnamefont {Page}},\ }\href {\doibase 10.1088/1361-6382/aae750}
  {\bibfield  {journal} {\bibinfo  {journal} {Class. Quantum Grav.}\ }\textbf
  {\bibinfo {volume} {35}},\ \bibinfo {pages} {224001} (\bibinfo {year}
  {2018})}\BibitemShut {NoStop}%
\bibitem [{\citenamefont {Sorkin}(1993)}]{sorkin1993impossible}%
  \BibitemOpen
  \bibfield  {author} {\bibinfo {author} {\bibfnamefont {R.~D.}\ \bibnamefont
  {Sorkin}},\ }in\ \href@noop {} {\emph {\bibinfo {booktitle} {Directions in
  general relativity: Proceedings of the 1993 International Symposium,
  Maryland}}},\ Vol.~\bibinfo {volume} {2}\ (\bibinfo {year} {1993})\ pp.\
  \bibinfo {pages} {293--305}\BibitemShut {NoStop}%
\bibitem [{\citenamefont {de~Ramón}\ \emph {et~al.}(2021)\citenamefont
  {de~Ramón}, \citenamefont {Papageorgiou},\ and\ \citenamefont
  {Martín-Martínez}}]{Pipolast}%
  \BibitemOpen
  \bibfield  {author} {\bibinfo {author} {\bibfnamefont {J.}~\bibnamefont
  {de~Ramón}}, \bibinfo {author} {\bibfnamefont {M.}~\bibnamefont
  {Papageorgiou}}, \ and\ \bibinfo {author} {\bibfnamefont {E.}~\bibnamefont
  {Martín-Martínez}},\ }\href@noop {} {\enquote {\bibinfo {title}
  {Relativistic causality in particle detector models: Faster-than-light
  signalling and "impossible measurements"},}\ } (\bibinfo {year} {2021}),\
  \Eprint {http://arxiv.org/abs/2102.03408} {arXiv (accepted in Phys. Rev.
  D):2102.03408 [quant-ph]} \BibitemShut {NoStop}%
\bibitem [{\citenamefont {Funai}\ and\ \citenamefont
  {Mart\'{\i}n-Mart\'{\i}nez}(2019)}]{NichoFaster}%
  \BibitemOpen
  \bibfield  {author} {\bibinfo {author} {\bibfnamefont {N.}~\bibnamefont
  {Funai}}\ and\ \bibinfo {author} {\bibfnamefont {E.}~\bibnamefont
  {Mart\'{\i}n-Mart\'{\i}nez}},\ }\href {\doibase 10.1103/PhysRevD.100.065021}
  {\bibfield  {journal} {\bibinfo  {journal} {Phys. Rev. D}\ }\textbf {\bibinfo
  {volume} {100}},\ \bibinfo {pages} {065021} (\bibinfo {year}
  {2019})}\BibitemShut {NoStop}%
\bibitem [{\citenamefont {Grebenkov}\ and\ \citenamefont
  {Nguyen}(2013)}]{review-laplacian}%
  \BibitemOpen
  \bibfield  {author} {\bibinfo {author} {\bibfnamefont {D.~S.}\ \bibnamefont
  {Grebenkov}}\ and\ \bibinfo {author} {\bibfnamefont {B.-T.}\ \bibnamefont
  {Nguyen}},\ }\href {\doibase 10.1137/120880173} {\bibfield  {journal}
  {\bibinfo  {journal} {SIAM Review}\ }\textbf {\bibinfo {volume} {55}},\
  \bibinfo {pages} {601} (\bibinfo {year} {2013})}\BibitemShut {NoStop}%
\bibitem [{\citenamefont {Robles}\ and\ \citenamefont
  {Claro}(2012)}]{Robles_2012}%
  \BibitemOpen
  \bibfield  {author} {\bibinfo {author} {\bibfnamefont {P.}~\bibnamefont
  {Robles}}\ and\ \bibinfo {author} {\bibfnamefont {F.}~\bibnamefont {Claro}},\
  }\href {\doibase 10.1088/0143-0807/33/5/1217} {\bibfield  {journal} {\bibinfo
   {journal} {European Journal of Physics}\ }\textbf {\bibinfo {volume} {33}},\
  \bibinfo {pages} {1217} (\bibinfo {year} {2012})}\BibitemShut {NoStop}%
\bibitem [{\citenamefont {Hilbert}\ and\ \citenamefont
  {Courant}(1968{\natexlab{a}})}]{courant-hilbert}%
  \BibitemOpen
  \bibfield  {author} {\bibinfo {author} {\bibfnamefont {D.}~\bibnamefont
  {Hilbert}}\ and\ \bibinfo {author} {\bibfnamefont {R.}~\bibnamefont
  {Courant}},\ }\href@noop {} {\emph {\bibinfo {title} {{Methoden der
  mathematischen Physik}}}},\ \bibinfo {series} {Heidelberger Tachenbuecher
  Band 30}, Vol.~\bibinfo {volume} {1}\ (\bibinfo  {publisher} {Springer},\
  \bibinfo {year} {1968})\BibitemShut {NoStop}%
\bibitem [{\citenamefont {Hilbert}\ and\ \citenamefont
  {Courant}(1968{\natexlab{b}})}]{courant-hilbert2}%
  \BibitemOpen
  \bibfield  {author} {\bibinfo {author} {\bibfnamefont {D.}~\bibnamefont
  {Hilbert}}\ and\ \bibinfo {author} {\bibfnamefont {R.}~\bibnamefont
  {Courant}},\ }\href@noop {} {\emph {\bibinfo {title} {{Methoden der
  mathematischen Physik}}}},\ \bibinfo {series} {Heidelberger Tachenbuecher
  Band 31}, Vol.~\bibinfo {volume} {2}\ (\bibinfo  {publisher} {Springer},\
  \bibinfo {year} {1968})\BibitemShut {NoStop}%
\bibitem [{\citenamefont {Kac}(1966)}]{kac-drum}%
  \BibitemOpen
  \bibfield  {author} {\bibinfo {author} {\bibfnamefont {M.}~\bibnamefont
  {Kac}},\ }\href@noop {} {\bibfield  {journal} {\bibinfo  {journal} {Am. Math.
  Mon.}\ }\textbf {\bibinfo {volume} {73}},\ \bibinfo {pages} {1} (\bibinfo
  {year} {1966})}\BibitemShut {NoStop}%
\bibitem [{\citenamefont {Aasen}\ \emph {et~al.}(2013)\citenamefont {Aasen},
  \citenamefont {Bhamre},\ and\ \citenamefont {Kempf}}]{kempf2013}%
  \BibitemOpen
  \bibfield  {author} {\bibinfo {author} {\bibfnamefont {D.}~\bibnamefont
  {Aasen}}, \bibinfo {author} {\bibfnamefont {T.}~\bibnamefont {Bhamre}}, \
  and\ \bibinfo {author} {\bibfnamefont {A.}~\bibnamefont {Kempf}},\ }\href
  {\doibase 10.1103/PhysRevLett.110.121301} {\bibfield  {journal} {\bibinfo
  {journal} {Phys. Rev. Lett.}\ }\textbf {\bibinfo {volume} {110}},\ \bibinfo
  {pages} {121301} (\bibinfo {year} {2013})}\BibitemShut {NoStop}%
\bibitem [{\citenamefont {Panine}\ and\ \citenamefont
  {Kempf}(2017)}]{kempf2017}%
  \BibitemOpen
  \bibfield  {author} {\bibinfo {author} {\bibfnamefont {M.}~\bibnamefont
  {Panine}}\ and\ \bibinfo {author} {\bibfnamefont {A.}~\bibnamefont {Kempf}},\
  }\href {\doibase 10.1142/S0219887817501572} {\bibfield  {journal} {\bibinfo
  {journal} {Int. J. Geom. Methods Mod. Phys}\ }\textbf {\bibinfo {volume}
  {14}},\ \bibinfo {pages} {1750157} (\bibinfo {year} {2017})}\BibitemShut
  {NoStop}%
\bibitem [{\citenamefont {Weyl}(1911)}]{Weyl1911}%
  \BibitemOpen
  \bibfield  {author} {\bibinfo {author} {\bibfnamefont {H.}~\bibnamefont
  {Weyl}},\ }\href@noop {} {\bibfield  {journal} {\bibinfo  {journal}
  {Nachrichten von der Gesellschaft der Wissenschaften zu Göttingen,
  Mathematisch-Physikalische Klasse}\ }\textbf {\bibinfo {volume} {1911}},\
  \bibinfo {pages} {110} (\bibinfo {year} {1911})}\BibitemShut {NoStop}%
\bibitem [{\citenamefont {Weyl}(1912)}]{Weyl1912}%
  \BibitemOpen
  \bibfield  {author} {\bibinfo {author} {\bibfnamefont {H.}~\bibnamefont
  {Weyl}},\ }\href {\doibase 10.1007/BF01456804} {\bibfield  {journal}
  {\bibinfo  {journal} {Math. Ann.}\ }\textbf {\bibinfo {volume} {71}},\
  \bibinfo {pages} {441} (\bibinfo {year} {1912})}\BibitemShut {NoStop}%
\bibitem [{\citenamefont {McCartin}(2003)}]{EquilateralDirichlet}%
  \BibitemOpen
  \bibfield  {author} {\bibinfo {author} {\bibfnamefont {B.~J.}\ \bibnamefont
  {McCartin}},\ }\href {http://www.jstor.org/stable/25054404} {\bibfield
  {journal} {\bibinfo  {journal} {SIAM Review}\ }\textbf {\bibinfo {volume}
  {45}},\ \bibinfo {pages} {267} (\bibinfo {year} {2003})}\BibitemShut
  {NoStop}%
\bibitem [{\citenamefont {J}(2002)}]{EquilateralNeumann}%
  \BibitemOpen
  \bibfield  {author} {\bibinfo {author} {\bibfnamefont {M.}~\bibnamefont
  {J}},\ }\href {\doibase 10.1080/1024123021000053664} {\bibfield  {journal}
  {\bibinfo  {journal} {Mathematical Problems in Engineering}\ }\textbf
  {\bibinfo {volume} {8}} (\bibinfo {year} {2002}),\
  10.1080/1024123021000053664}\BibitemShut {NoStop}%
\bibitem [{\citenamefont {McCartin}(2004)}]{EquilateralRobin}%
  \BibitemOpen
  \bibfield  {author} {\bibinfo {author} {\bibfnamefont {B.}~\bibnamefont
  {McCartin}},\ }\href {\doibase 10.1155/S0161171204306125} {\bibfield
  {journal} {\bibinfo  {journal} {International Journal of Mathematics and
  Mathematical Sciences}\ }\textbf {\bibinfo {volume} {2004}} (\bibinfo {year}
  {2004}),\ 10.1155/S0161171204306125}\BibitemShut {NoStop}%
\bibitem [{\citenamefont {Hale}(2005)}]{hale}%
  \BibitemOpen
  \bibfield  {author} {\bibinfo {author} {\bibfnamefont {J.}~\bibnamefont
  {Hale}},\ }in\ \href {\doibase 10.1016/B978-044451861-3/50003-3} {\emph
  {\bibinfo {booktitle} {10 Mathematical Essays on Approximation in Analysis
  and Topology}}},\ \bibinfo {editor} {edited by\ \bibinfo {editor}
  {\bibfnamefont {J.}~\bibnamefont {Ferrera}}, \bibinfo {editor} {\bibfnamefont
  {J.}~\bibnamefont {López-Gómez}}, \ and\ \bibinfo {editor} {\bibfnamefont
  {F.}~\bibnamefont {{Ruiz del Portal}}}}\ (\bibinfo  {publisher} {Elsevier
  Science},\ \bibinfo {address} {Amsterdam},\ \bibinfo {year} {2005})\ pp.\
  \bibinfo {pages} {95--123}\BibitemShut {NoStop}%
\bibitem [{\citenamefont {Burenkov}\ and\ \citenamefont
  {Davies}(2002)}]{davies2002}%
  \BibitemOpen
  \bibfield  {author} {\bibinfo {author} {\bibfnamefont {V.}~\bibnamefont
  {Burenkov}}\ and\ \bibinfo {author} {\bibfnamefont {E.}~\bibnamefont
  {Davies}},\ }\href {\doibase 10.1016/S0022-0396(02)00033-5} {\bibfield
  {journal} {\bibinfo  {journal} {J. Diff. Eq.}\ }\textbf {\bibinfo {volume}
  {186}},\ \bibinfo {pages} {485} (\bibinfo {year} {2002})}\BibitemShut
  {NoStop}%
\bibitem [{\citenamefont {Hempel}\ \emph {et~al.}(1991)\citenamefont {Hempel},
  \citenamefont {Seco},\ and\ \citenamefont {Simon}}]{hempel}%
  \BibitemOpen
  \bibfield  {author} {\bibinfo {author} {\bibfnamefont {R.}~\bibnamefont
  {Hempel}}, \bibinfo {author} {\bibfnamefont {L.~A.}\ \bibnamefont {Seco}}, \
  and\ \bibinfo {author} {\bibfnamefont {B.}~\bibnamefont {Simon}},\ }\href
  {\doibase 10.1016/0022-1236(91)90130-W} {\bibfield  {journal} {\bibinfo
  {journal} {J. Funct. Anal.}\ }\textbf {\bibinfo {volume} {102}},\ \bibinfo
  {pages} {448} (\bibinfo {year} {1991})}\BibitemShut {NoStop}%
\bibitem [{\citenamefont {Louko}\ and\ \citenamefont
  {Satz}(2006)}]{Louko_2006}%
  \BibitemOpen
  \bibfield  {author} {\bibinfo {author} {\bibfnamefont {J.}~\bibnamefont
  {Louko}}\ and\ \bibinfo {author} {\bibfnamefont {A.}~\bibnamefont {Satz}},\
  }\href {\doibase 10.1088/0264-9381/23/22/015} {\bibfield  {journal} {\bibinfo
   {journal} {Classical and Quantum Gravity}\ }\textbf {\bibinfo {volume}
  {23}},\ \bibinfo {pages} {6321} (\bibinfo {year} {2006})}\BibitemShut
  {NoStop}%
\end{thebibliography}%

\end{document}